\documentclass[twocolumn,showpacs,amsmath,amssymb,superscriptaddress]{revtex4-1} %,groupedaddress
\usepackage{dcolumn,braket,color,amsmath,footnote,hyperref,bm}
\usepackage{graphicx,bm,url,longtable}
\usepackage{tabularx,multirow,braket}
\usepackage{fancybox}
\usepackage{tikz}
\usetikzlibrary{matrix}

\newcommand\+{\dagger}
\newcommand\pv{${\mathit pv}$}

\begin{document}

\title{Pairing vibrations in the interacting boson model based on density functional theory}

\author{K.~Nomura}
\affiliation{Department of Physics, Faculty of Science, University of
Zagreb, HR-10000 Zagreb, Croatia}
\email{knomura@phy.hr}
\author{D.~Vretenar}
\affiliation{Department of Physics, Faculty of Science, University of
Zagreb, HR-10000 Zagreb, Croatia}
 \affiliation{ State Key Laboratory of Nuclear Physics and Technology, School of Physics, Peking University, Beijing 100871, China}
\author{Z.~P.~Li}
\affiliation{School of Physical Science and Technology, Southwest
University, Chongqing 400715, China}
%\affiliation{Department of Physics, Faculty of Science, University of
%Zagreb, HR-10000 Zagreb, Croatia}

\author{J.~Xiang}
\affiliation{School of Physics and Electronic, Qiannan Normal University
for Nationalities, Duyun 558000, China}
\affiliation{School of Physical Science and Technology, Southwest
University, Chongqing 400715, China}

\date{\today}

\begin{abstract}
We propose a method to incorporate the coupling between shape and pairing 
collective degrees of freedom in the framework of the interacting boson  model (IBM), 
based on the nuclear density functional theory. 
To account for pairing vibrations, a boson-number non-conserving IBM
 Hamiltonian is introduced. The Hamiltonian is constructed by using
 solutions of  self-consistent mean-field calculations
based on a universal energy density functional and pairing force, with 
 constraints on the axially-symmetric quadrupole and pairing intrinsic deformations. 
By mapping the resulting quadrupole-pairing potential energy surface
 onto the expectation value of the bosonic Hamiltonian in the boson
 condensate state, the strength parameters of the boson Hamiltonian are
 determined. An illustrative calculation is performed for
 $^{122}$Xe, and the method is further explored in a more systematic study of 
 rare-earth $N=92$ isotones. 
The inclusion of the dynamical pairing degree of
 freedom significantly lowers the energies of bands based on excited $0^+$ states. 
The results are in quantitative agreement with spectroscopic data,
 and are consistent with those obtained using the collective Hamiltonian
 approach. 
\end{abstract}

\maketitle

\section{Introduction}

Pairing correlations are among the most prominent features of the nuclear 
many-body system \cite{bohr1958,BM,RS,brink-broglia} and, to a large 
extent, determine the structure of low-energy nuclear spectra. 
%The pairs are often broken up into single nucleons (quasiparticles) and,
%in such cases, interplay between collective 
%and single-particle degrees of freedom becomes even more significant. 
Pairing vibrations \cite{bohr1964,bes1966,broglia1973,brink-broglia}, in particular, 
play an important role in fundamental processes
such as neutrinoless $\beta\beta$ decay \cite{vaquero2013}, and
spontaneous fission 
\cite{giuliani2014,zhao2016,rayner2018,rayner2020}. 
The relevance of pairing vibrations in structure phenomena 
has been investigated using a variety of 
nuclear models. Here we particularly refer to theoretical studies
since the early 2000's, that have used the geometrical collective Hamiltonian 
\cite{sieja2004,pomorski2007,prochniak2007,xiang2020}, the 
time-dependent Hartree-Fock-Bogoliubov 
approaches \cite{avez2008}, the nuclear shell model \cite{heusler2015},
the quasiparticle random-phase approximation
\cite{khan2009,shinomiya2011}, and the generator coordinate methods
(GCM) \cite{vaquero2011,vaquero2013}. 

Nuclear density functional theory (DFT) is at present the most
reliable framework for the description of low-energy structure of medium-heavy 
and heavy nuclei. Both the relativistic
\cite{vretenar2005,niksic2011,meng2016} and nonrelativistic
\cite{bender2003,erler2011,robledo2019} energy density functionals
(EDFs) have been successfully implemented the self-consistent mean-field (SCMF)
studies of static and dynamical properties of finite nuclei. 
Within this framework, the calculation of excitation spectra requires the 
restoration of broken symmetries and configuration mixing, e.g., using the
generator coordinate method (GCM) \cite{RS}. 
However, when multiple collective coordinates need to be taken into
account, this type of calculation becomes computationally excessive. 
In the recent work of Ref.~\cite{xiang2020}, the coupling between shape and 
pairing degrees of freedom has been considered using a quadrupole plus 
pairing collective Hamiltonian based on the relativistic mean-field plus
Bardeen-Cooper-Schrieffer (RMF+BCS) scheme. It has been shown that 
the inclusion of the pairing degree of freedom 
significantly improves the description of low-lying $0^+$ states 
in rare-earth nuclei. The current implementation of this approach is,
however, restricted to axially-symmetric shapes.

Nuclear spectroscopy is also studied with a theoretical method
that consists in mapping the solutions of the DFT SCMF calculation onto
the interacting-boson Hamiltonian \cite{nomura2008,nomura2010}. 
The interacting boson model (IBM) \cite{arima1975,IBM}, originally
introduced by Arima and Iachello, is a model in which correlated pairs of
valence nucleons with spin and parity $0^+$ and $2^+$ are approximated
by effective bosonic degrees of freedom ($s$ and $d$ bosons,
respectively) \cite{OAI,IBM}. 
In the DFT-to-IBM mapping procedure of Ref.~\cite{nomura2008}, the strength
parameters of the IBM Hamiltonian are completely determined by mapping a 
SCMF potential energy surface (PES), obtained from constrained SCMF 
calculations with a choice of the EDF and pairing force, onto the
expectation value of the Hamiltonian in the boson coherent state
\cite{ginocchio1980}. The method has been successfully applied 
in studies of a variety of interesting nuclear structure 
phenomena, such as shape coexistence \cite{nomura2016zr,nomura2016sc}, 
octupole collective excitations 
\cite{nomura2013oct,nomura2014,nomura2018oct,nomura2020oct}, 
quantum phase transitions in odd-mass and odd-odd nuclei
\cite{nomura2016odd,nomura2020cs,nomura2020zr}, 
and $\beta$ decay \cite{nomura2020beta-1,nomura2020beta-2}.

Considering the microscopic basis of the IBM in which the bosons represent
valence nucleon pairs \cite{OAIT,OAI,IBM}, one might attempt to 
implement also pairing vibrational modes in the IBM. 
In Refs.~\cite{vanisacker1982,hasegawa1985,kaup1988a,kaup1988b} additional
monopole boson degrees of freedom, different from the standard 
$s$ boson, were introduced in the IBM to reproduce
low-lying excited $0^+$ energies. 
Because of the inclusion of new building blocks, however, the number of free 
parameters increases in such an approach. 
Except for the references above, very little progress has been made in 
explicitly including pairing vibrations in the IBM framework.

In this work, we develop a method to incorporate both shape and pairing
vibrations in the IBM. To account for the pairing degree of freedom, we introduce a
version of the IBM (denoted hereafter by \pv-IBM) in which the number of
bosons is not conserved but is allowed to change by one. 
Subsequently the boson space consists of three subspaces that differ in
boson number by one. The three subspaces are mixed by a specific
monopole pair transfer operator. 
The strength parameters of the \pv-IBM Hamiltonian are completely
determined by the mapping of the SCMF $(\beta,\alpha)$ potential energy surface,
obtained from RMF+BCS calculations, onto the bosonic counterpart. 
We demonstrate that the inclusion of dynamical pairing in the IBM framework 
significantly lowers the energies of excited $0^+$ states, in very good agreement with data.

The paper is organised as follows. In Sec.~\ref{sec:SCMF} we briefly
review the underlying SCMF calculations. In Sec.~\ref{sec:IBMpv} the
\pv-IBM model is introduced, and a method for mapping the SCMF onto
bosonic deformation energy surfaces is described. The model is 
illustrated using as an example the excitation spectrum of the nucleus $^{122}$Xe in Sec.~\ref{sec:xe122}. 
In Sec.~\ref{sec:N92} the newly developed method is further explored in 
a study of low-energy $K^\pi=0^+$ bands in 
four axially-symmetric $N=92$ rare-earth isotones. 
Section~\ref{sec:summary} presents a summary of the main
results and an outlook for future study.

\section{Quadrupole-and-pairing constrained SCMF calculation\label{sec:SCMF}}

\begin{figure}[htb!]
\begin{center}
\includegraphics[width=0.7\linewidth]{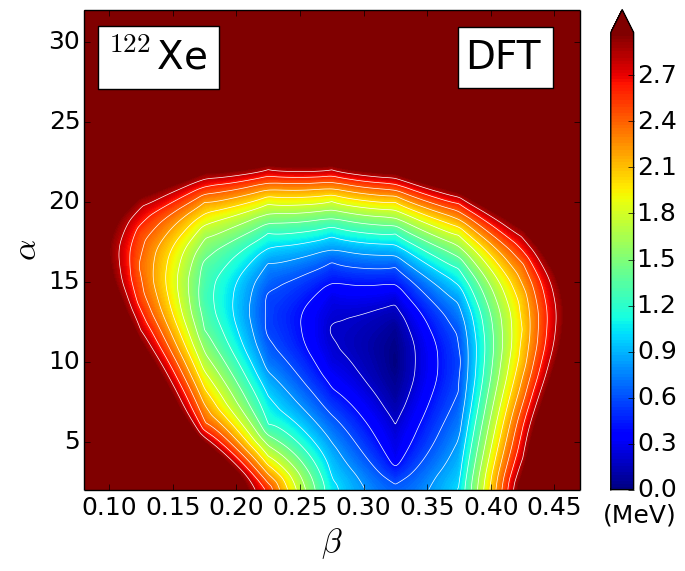} 
\caption{(Color online) The potential energy surface (PES) of $^{122}$Xe in the $(\beta,\alpha)$ plane, calculated by the constrained RMF+BCS with the PC-PK1 energy density functional and separable pairing interaction. All energies (in MeV) in the PES are normalized with respect to the binding energy of the absolute minimum. The contours join points on the surface with the same energies, and energy difference between the neighbouring contours is 200 keV.}
\label{fig:pes-dft}
\end{center}
\end{figure}

In a first step, constrained self-consistent mean-field (SCMF)
calculations are performed within the framework of the relativistic
mean-field plus BCS (RMF+BCS) model. In the present study, the
particle-hole channel of the effective inter-nucleon interaction is determined by 
the universal energy density functional PC-PK1 \cite{PCPK1}, while the particle-particle channel is
modeled in the BCS approximation using a separable pairing force
\cite{tian2009}. A more detailed description of the RMF+BCS framework combined
with the separable pairing force can be found
in Ref.~\cite{xiang2012}. The constraints imposed in the SCMF calculation are on the 
expectation values of axial quadrupole $\hat Q_{20}$ and monopole pairing $\hat P_0$ operators. The
quadrupole operator $\hat Q_{20}$ is defined as $\hat Q_{20} = 2z^2 - x^2
- y^2$, and its expectation value corresponds to the dimensionless axial
deformation parameter $\beta$: 
\begin{align}
\label{eq:beta-f}
 \beta = \frac{\sqrt{5\pi}}{3r_0^2A^{5/3}}\braket{\hat Q_{20}},
\end{align}
with $r_0=1.2$ fm. If one does not consider ``pairing rotations'', that is, quasirotational bands that
correspond to ground states of neighboring even-even nuclei, the monopole pairing operator 
takes a simple form: 
\begin{align}
\label{eq:pair-f}
\hat P_0 = \frac{1}{2}\sum_{k>0}(c_kc_{\bar{k}}+c^\+_{\bar k}c^\+_{k})\;,
\end{align}
where $k$ and ${\bar{k}}$ denote the single-nucleon and the corresponding time-reversed states, respectively. 
$c^\+_{k}$ and $c_{k}$ are the single-nucleon creation and annihilation operators. 
The expectation value of the pairing operator in a BCS state
\begin{align}
%\ket{\alpha\phi}=e^{iN\phi}\Pi_{k>0}(u_k + v_k e^{-2i\phi}c_kc_{\bar{k}}c^\+_{\bar k}c^\+_{k})\ket{0} 
\ket{\alpha}=\prod_{k>0}(u_k + v_k c^\+_{\bar k}c^\+_{k})\ket{0}, 
\end{align}
corresponds to the intrinsic pairing deformation parameter $\alpha$,
\begin{align}
\label{eq:alph-f}
\alpha = \braket{\alpha|\hat P_0|\alpha} = \sum_{\tau}\sum_{k>0}u_k^{\tau} v_k^{\tau}, 
\end{align}
which can be related to the pairing gap $\Delta$. 
The sum runs over both proton $\tau=\pi$ and neutron $\tau=\nu$ single-particle states. 
The quadrupole shape deformation Eqs.~(\ref{eq:beta-f}) and pairing deformation (\ref{eq:alph-f}) 
represent the collective coordinates for constrained SCMF calculations \cite{xiang2020}. 
As an example, in Fig.~\ref{fig:pes-dft} we display the SCMF deformation 
energy surface for the nucleus $^{122}$Xe in the plane of the axial
quadrupole $\beta$ and pairing $\alpha$ deformation variables. 
The global minimum is found at $\beta\approx 0.32$ and 
$\alpha\approx 10$, and we note that the energy surface is rather soft with respect to the 
pairing variable $\alpha$.

%-----------------------------------------------------------
%
%	SCMF QP-PES
%
%-----------------------------------------------------------

\section{Pairing vibrations in the IBM\label{sec:IBMpv}}

\subsection{The Hamiltonian}

In the next step we introduce a model that relates the SCMF
($\beta, \alpha$) potential energy surface (PES) to an equivalent system of interacting bosons. 
The boson space comprises monopole $s$ and
quadrupole $d$ bosons, which represent correlated $L=0^+$ and $2^+$
pairs of valence nucleons \cite{OAIT,OAI,IBM}. 
In the conventional IBM, the number of bosons, denoted
as $n$, is conserved for a given nucleus, i.e., $n=n_s + n_d$, where
$n_s$ and $n_d$ stand for the $s$ and $d$ boson number, respectively. 
The boson number is equal to half the number of 
valence nucleons counted from the nearest closed shells and, in the illustrative case 
$^{122}$Xe, the boson core nucleus is $^{132}$Sn and hence $n=9$. 
We do not distinguish between neutron and proton degrees freedom in the boson space. 
%The monopole pair operator in the fermion system is defined as in Eq.~(\ref{eq:pair-f}). 
Considering the underlying microscopic structure, the monopole pair transfer
operator in the bosonic system should be expressed, to a good 
approximation, in terms of the $s$ boson degree of freedom, i.e., $\hat
P \propto s^\+ + s$. 
Hence the $s$ boson is expected to be the most relevant for a description of the pairing vibration mode.

To take explicitly into account the pairing vibration mode, the boson
configuration space is extended in such a way that the total number of
bosons is no longer conserved, but is allowed to change in the boson number by
one, that is, $8\leqslant n \leqslant 10$ for the illustrative case of $^{122}$Xe. The following IBM Hamiltonian
is employed: 
%\begin{widetext}
\begin{align}
\label{eq:ham}
\hat H=
& \sum_{n}\hat{\mathcal P}_n
(\epsilon_s^0 \hat n_s + \epsilon_d^0\hat n_d + \kappa\hat Q\cdot\hat Q + \kappa'\hat L\cdot\hat L)
\hat{\mathcal P}_n \nonumber \\
& + \sum_{n\neq n'}\hat{\mathcal P}_{n} t_s(s^\dagger + s) \hat{\mathcal P}_{n'}
\end{align}
%\end{widetext}
where $\hat{\mathcal P}_n$ is the projection operator onto the subspace $[n]$. 
The parameters for the Hamiltonian could differ between different 
configuration spaces, but here the same parameters are used 
for the three configurations. 
Therefore, for brevity, in the following the operator $\hat{\mathcal P}_n$ will be omitted, 
unless otherwise specified. 
The first and second terms in Eq.~(\ref{eq:ham}) are the $s$ and $d$ boson-number operators 
with $\hat n_s = s^\dagger\cdot s$ and $\hat n_d = d^\dagger\cdot\tilde d$. 
$\epsilon_s^0$ and $\epsilon_d^0$ are absolute values of the 
single $s$ and $d$ boson energies. 
The third term is the quadrupole-quadrupole interaction with the 
boson quadrupole operator $\hat Q=s^\dagger\tilde d + d^\dagger s + \chi (d^\+\times\tilde d)^{(2)}$. 
The fourth term, with the boson angular momentum operator 
$\hat L=\sqrt{10}(d^\+\times\tilde d)^{(1)}$, makes a significant
contribution to the moments of inertia of the $K^\pi=0^+$ bands.  
The last term with strength $t_s$ in the above Hamiltonian  
represents the one $s$-boson (monopole pair) transfer operator. 
It is the boson-number non-conserving term, and thus mixes the subspaces 
$[n-1]$, $[n]$, and $[n+1]$. 
For later convenience, and since the total boson number operator 
is given as $\hat n=\hat n_s + \hat n_d$, the above Hamiltonian is
rewritten in the form: 
\begin{align}
\label{eq:ham2}
\hat H=\epsilon_s^0 \hat n + \epsilon_d\hat n_d + 
\kappa\hat Q\cdot\hat Q + \kappa'\hat L\cdot\hat L + t_s(s^\dagger + s)
\end{align}
where $\epsilon_d$ is the $d$-boson energy relative to the $s$ boson
one, i.e., $\epsilon_d = \epsilon^0_d - \epsilon^0_s$. 
The first term $\epsilon_s^0 \hat n$ does not contribute to the relative 
excitation spectra, and is thus neglected in most IBM calculations. 
In the present framework, however, since we allow for the boson number 
to vary, this global term is expected to play an important role, especially 
for excitation energies of the $0^+$ states.

The Hamiltonian Eq.~(\ref{eq:ham2}) is diagonalized 
in the following $M$-scheme basis with $M=0$, expressed as a
direct sum of the bases for the three configurations: 
\begin{align}
\label{eq:basis}
\ket{\Phi} = 
[\ket{(sd)^{n-1}}\oplus\ket{(sd)^{n}}\oplus\ket{(sd)^{n+1} }]_{M=0},
%[N_\mathrm{B}-1]\oplus[N_\mathrm{B}]\oplus[N_\mathrm{B}+1]
\end{align}
where $M$ denotes the $z$-projection of the total angular momentum $I$. 
The value of $I$ for a given eigenstate is identified by calculating the
expectation value of the angular momentum operator squared, which should
give the eigenvalue $I(I+1)$. 

The present computational scheme is formally similar to IBM 
configuration-mixing calculations that describe the
phenomenon of shape coexistence \cite{nomura2016sc}. 
In the conventional configuration-mixing IBM framework, 
several different boson Hamiltonians are allowed to mix \cite{duval1981}. 
Each of these independent (unperturbed) Hamiltonians is associated with 
a $2m$-particle-$2m$-hole ($m\in \mathbb{Z}$) excitation from a given
major shell to the next and, since in the IBM there is no distinction between particles and holes, 
differ in boson number by two. The
configuration-mixing IBM thus does not conserve the boson number, similar to 
the present case. Here, however, the model space comprises a single
major shell, and the boson number conservation is violated not by the
contribution from next major shell (i.e., pair transfer across the shell
closure), but by pairing vibrations.

\subsection{The boson condensate}

The IBM analogue of the $(\beta,\alpha)$ PES is formulated analytically
by taking the expectation value of the Hamiltonian of  
Eq.~(\ref{eq:ham}) in the boson coherent state $\ket{\Psi(\vec\alpha)}$
\cite{ginocchio1980,bohr1980,hatch1982}: 
\begin{align}
\label{eq:coherent}
\ket{\Psi({\vec\alpha})}
=\ket{\Psi(n-1,{\vec\alpha})}\oplus 
\ket{\Psi(n,{\vec\alpha})}\oplus
\ket{\Psi(n+1,{\vec\alpha})},
\end{align}
where $\vec{\alpha}$ represents variational parameters. 
Since here the IBM model space comprises three different boson-number configurations,
the above trial wave function is expressed as a direct sum of 
three independent coherent states. 
Each of them is given by
\begin{align}
\ket{\Psi(n,{\vec\alpha})}=\frac{1}{\sqrt{n!}}(b_c^\+)^n\ket{0}, 
\end{align}
and the condensate boson 
$b_c$ is defined as
\begin{align}
\label{eq:bc}
b_c = (\alpha_0^2+\alpha_2^2)^{-1/2}(\alpha_0 s + \alpha_2 d_0), 
\end{align}
where the amplitudes $\alpha_{0}$ and $\alpha_2$ should be related to
the pairing deformation  $\alpha$ and the axial deformation parameter
$\beta$ in the SCMF calculation, respectively. 
The variable $\alpha_2$ can be considered as the shape 
deformation parameter in the collective model:  
\begin{align}
\label{eq:beta}
\alpha_2 = \bar\beta, 
\end{align}
where $\bar\beta$ is the IBM analog of the 
axially symmetric SCMF deformation parameter. 
%In order to better associate with the SCMF $(\beta,\alpha)$ energy surface, 
We propose to perform the following coordinate transformation for the
variable $\alpha_0$:
\begin{align}
\label{eq:cosh}
\alpha_0  = \cosh{(\bar\alpha-\bar\alpha_\mathrm{min})}. 
\end{align}
The new coordinate $\bar\alpha$ is equivalent to the pairing deformation
$\alpha$. $\bar\alpha_\mathrm{min}$ stands for the
$\bar\alpha$ value corresponding to the global minimum on the IBM PES. 
We assume the following relations that relate the amplitudes $\bar\beta$ and $\bar\alpha$ 
in the boson system to the $\beta$ and $\alpha$ coordinates of the SCMF model: 
\begin{align}
\bar\beta = C_\beta\beta, \quad \bar\alpha =
 C_\alpha\alpha. 
\end{align} 
The dimensionless coefficients of proportionality $C_\beta$ and $C_\alpha$ are 
additional scale parameters determined by the mapping.

Since our model space comprises three subspaces with different number of bosons, the PES of the boson system 
%with the $\bar\beta$ and $\bar\alpha$ degrees of freedom 
is expressed in a matrix form \cite{frank2004}: 
\begin{align}
\label{eq:pes-mat}
\Biggr(
\begin{array}{ccc}
E_{n-1,n-1}(\bar\alpha,\bar\beta) & E_{n-1,n}(\bar\alpha,\bar\beta) & 0 \\
E_{n,n-1}(\bar\alpha,\bar\beta) & E_{n,n}(\bar\alpha,\bar\beta) & E_{n,n+1}(\bar\alpha,\bar\beta) \\
0 & E_{n+1,n}(\bar\alpha,\bar\beta) & E_{n+1,n+1}(\bar\alpha,\bar\beta) 
\end{array}
\Biggl)
\end{align}
In the limit in which boson number is conserved, only the 
diagonal element $E_{n,n}$ is considered. 
The energy-surface matrix of Eq.~(\ref{eq:pes-mat}) is diagonalized 
at each point on the surface ($\bar\beta,\bar\alpha$), resulting in 
three energy surfaces \cite{frank2004}. The usual procedure in most IBM calculations with configuration mixing 
is to retain only the lowest energy eigenvalue at each deformation.

The analytical expressions for the 
diagonal and non-diagonal elements of the matrix Eq.~(\ref{eq:pes-mat}) are 
obtained by calculating expectation values of the Hamiltonian 
Eq.~(\ref{eq:ham2}) in the coherent state Eq.~(\ref{eq:bc}), with the 
amplitudes defined in Eqs.~(\ref{eq:beta}) and (\ref{eq:cosh}). 
The right-hand side of Eq.~(\ref{eq:cosh}) is Taylor expanded: 
$\cosh{\bar\alpha'} = 1 + \bar\alpha'^2/2 + 
\mathcal{O}(\bar\alpha'^4)$, and thus  
$\cosh^2{\bar\alpha'} = 1 + \bar\alpha'^2 +
\mathcal{O}(\bar\alpha'^4)$, 
where $\bar\alpha'\equiv\bar\alpha-\bar\alpha_\mathrm{min}$. 
Terms of the order of $\mathcal{O}(\bar\alpha'^4)$ and higher are
hereafter neglected. The resulting analytical expressions for the matrix
elements in Eq.~(\ref{eq:pes-mat}) read: 
%\begin{widetext}
\begin{align}
\label{eq:pes-diag}
&E_{n,n}(\bar\alpha,\bar\beta)
=
\epsilon^0_s n 
+ \frac{n[5\kappa+(\epsilon_d+6\kappa'+\kappa(1+\chi^2))\bar\beta^2]}{1+\bar\alpha'^2+\bar\beta^2}
 \nonumber \\
& + \frac{\kappa n(n-1)}{(1+\bar\alpha'^2+\bar\beta^2)^2}
\Biggl[
4(1+\bar\alpha'^2)\bar\beta^2 
- 4\sqrt{\frac{2}{7}}\chi\bar\beta^3
+\frac{2}{7}\chi^2\bar\beta^4
\Biggr],
\end{align}
for the diagonal elements, and  
\begin{align}
\label{eq:pes-nondiag}
&E_{n,n'}(\bar\alpha,\bar\beta)=E_{n',n}(\bar\alpha,\bar\beta)
=t_s \frac{2\sqrt{n+1}}{\sqrt{1+\bar\alpha'^2 + \bar\beta^2}},
%\quad (n>n'), 
\end{align}
%\end{widetext}
for the non-diagonal elements with $n>n'$. 
The term proportional to $\bar\alpha'^2\bar\beta^3$ in the
numerator of the third term of Eq.~(\ref{eq:pes-diag}), and the term
quadratic in $\bar\alpha'$ in the numerator of  
Eq.~(\ref{eq:pes-nondiag}) are neglected. 

The functional forms in Eqs.~(\ref{eq:pes-diag})
and (\ref{eq:pes-nondiag}), in particular the norm factor
$\mathcal{N}=1+\bar\alpha'^2+\bar\beta^2$ that depends quadratically on $\bar\alpha$, most
effectively produce an $\alpha$-deformed equilibrium state that is consistent with the SCMF
PES. The form of the norm factor ${\mathcal N}$ ensures that 
no divergence occurs at $\bar\beta\approx 0$ and 
$\bar\alpha'\approx 0$. In the limit $\bar\alpha\to \bar\alpha_{\rm min}
(=C_\alpha\alpha_\mathrm{min})$, the
expression for $E_{n,n}(\bar\alpha,\bar\beta)$ reduces to the one
used in standard $sd$-IBM calculations 
\cite{ginocchio1980,IBM}.

%-----------------------------------------------------------
%
%	IBM QP-PES
%
%-----------------------------------------------------------
\begin{figure}[htb!]
\begin{center}
\includegraphics[width=0.7\linewidth]{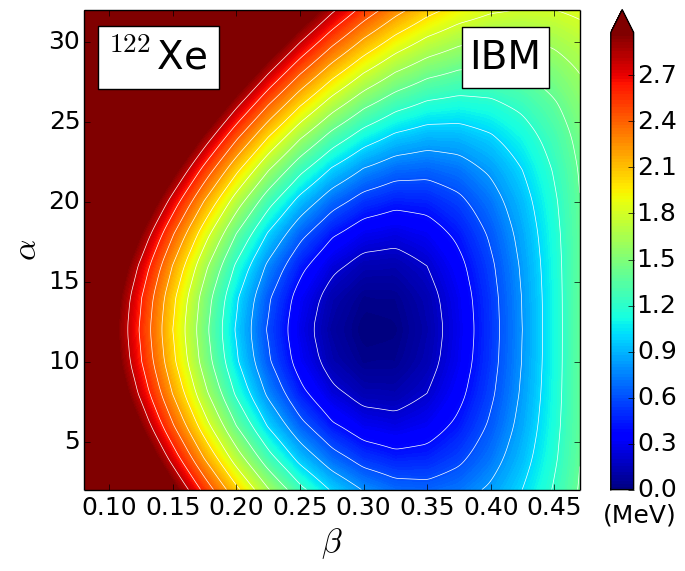} 
\caption{(Color online) Same as in the caption to Fig.~\ref{fig:pes-dft} 
but for the IBM energy surface.} 
\label{fig:pes-ibm}
\end{center}
\end{figure}

\subsection{Mapping the boson Hamiltonian}

The \pv-IBM Hamiltonian in Eq.~(\ref{eq:ham2}) is constructed in the
following steps: 
\begin{enumerate}

\item The strength parameters that appear in the boson number conserving part of
      the Hamiltonian: $\epsilon_d$, $\kappa$, and $\chi$, as well as
      the scale factor $C_\beta$, are determined so that the diagonal matrix element 
$E_{n,n}(\bar\alpha,\bar\beta)$ reproduces the SCMF PES at
      $\alpha=\alpha_{\rm min}$.  

\item The strength $\kappa'$ of the rotational $\hat L\cdot\hat L$ term 
is determined separately so that the bosonic moment of
      inertia calculated in the intrinsic frame \cite{schaaser1986} at
      the global minimum, should equal the 
Inglis-Belyaev \cite{inglis1956,belyaev1961} moment of inertia at the
      corresponding configuration on the SCMF energy surface. The details of this 
      procedure can be found in
      Ref.~\cite{nomura2011rot}.  

\item The $s$ boson energy $\epsilon^0_s$ and mixing strength $t_s$, 
as well as the scale factor $C_\alpha$, 
are determined in the $(\bar\beta, \bar\alpha)$ plane so that the lowest
      eigenvalue of the energy surface matrix Eq.~(\ref{eq:pes-mat})
      reproduces the topology of the SCMF PES in the neighbourhood
      of the equilibrium minimum. 

\end{enumerate}

%-----------------------------------------------------------------------
%
%	PARAMETERS
%
%-----------------------------------------------------------------------
\begin{table}[!htb]
\begin{center}
\caption{\label{tab:parameter} 
Strength parameters of the boson Hamiltonian Eq.~(\ref{eq:ham2}) 
determined by mapping the SCMF $(\beta, \alpha)$ energy surface to the boson space. The
 parameters $\chi$, $C_\beta$, and $C_\alpha$ are dimensionless, while
 the others are in units of MeV.}
 \begin{ruledtabular}
 \begin{tabular}{cccccccc}
  $\epsilon_s^0$ & $\epsilon_d$ & $\kappa$ & $\chi$ & $\kappa'$
  &  $t_s$  & $C_\beta$ & $C_\alpha$ \\
\hline
2.19 & 0.611 & $-0.102$ & $-0.4$ & $-0.029$ & 0.18 & 2.75 & 0.045 \\
 \end{tabular}
 \end{ruledtabular}
\end{center} 
\end{table}

The values of the resulting parameters of the IBM Hamiltonian are listed in
Table~\ref{tab:parameter}, and the corresponding IBM PES is shown in
Fig.~\ref{fig:pes-ibm}. Consistent with the SCMF PES, the equilibrium minimum of the 
IBM PES is found at $\beta\approx 0.32$ and
$\alpha\approx 10$. 
The potential energy surfaces exhibit a similar topography
except for the fact that, away from the global minimum, the IBM surface
tends to be softer than the DFT one obtained using the constrained SCMF method.  
This is a common characteristic of the IBM \cite{nomura2008} that arises because of 
the more restricted boson model space as compared to the SCMF approach based on 
the Kohn-Sham DFT. The former is built only from the valence nucleons, while the latter model 
space contains all nucleons. Therefore, the boson Hamiltonian parameters are determined 
by the mapping procedure that is carried out in the
neighbourhood of the global minimum, as this region is most relevant for low-energy excitations. The Hamiltonian 
Eq.~(\ref{eq:ham2}) is diagonalised in the $M$-scheme basis of
Eq.~(\ref{eq:basis}).

%-----------------------------------------------------------
%
%	UNPERTURBED V.S. PERTURBED
%
%-----------------------------------------------------------
\begin{figure}[htb!]
\begin{center}
\includegraphics[width=\linewidth]{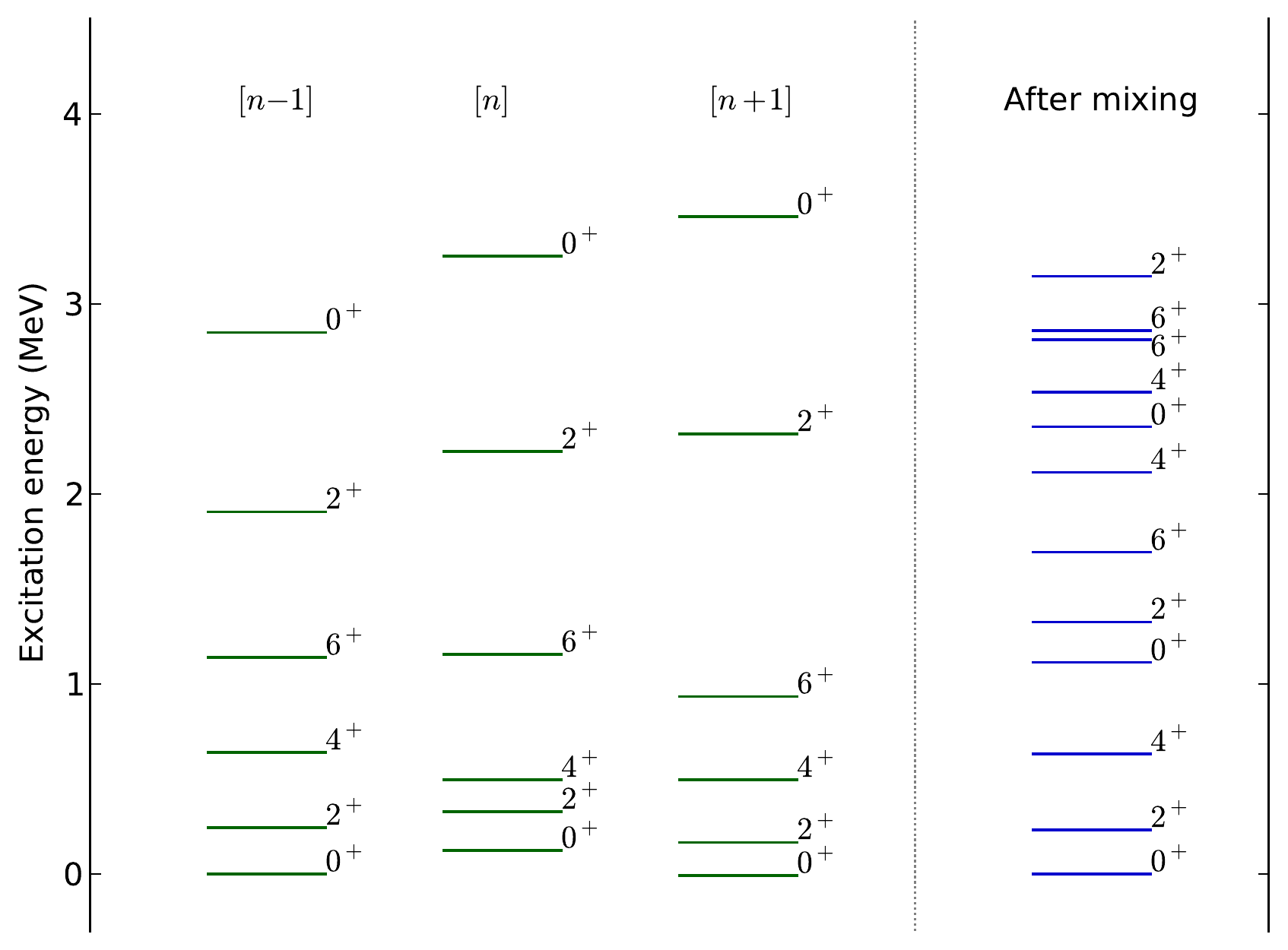}
\caption{(Color online) Excitation spectra for the unperturbed boson configurations 
$[n-1]$, $[n]$ and $[n+1]$, and the final one obtained after mixing the three
 configurations.} 
\label{fig:unptb}
\end{center}
\end{figure}

%-----------------------------------------------------------
%
%	Spectra
%
%-----------------------------------------------------------
\begin{figure*}[htb!]
\begin{center}
\includegraphics[width=0.7\linewidth]{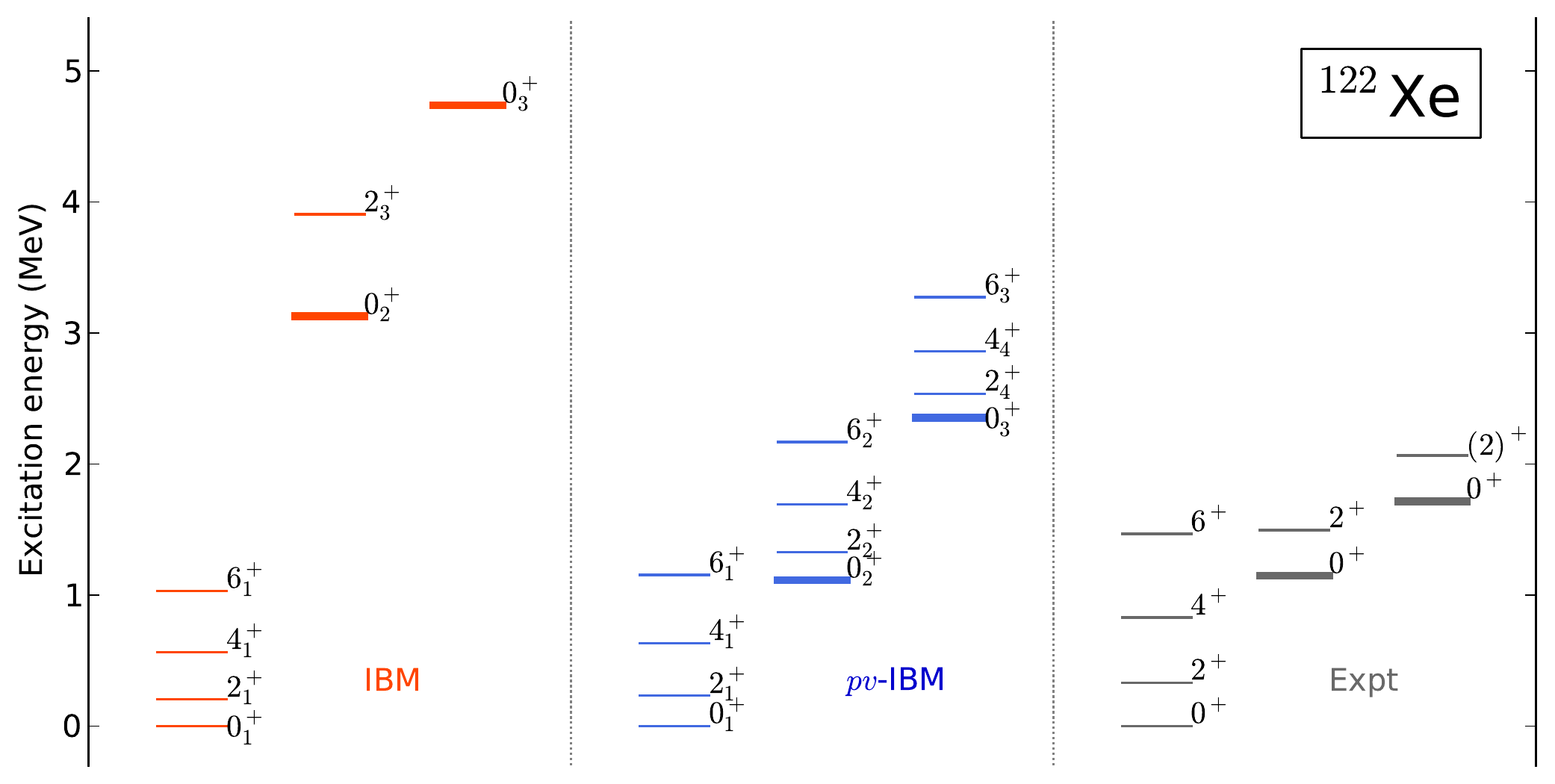}
\caption{(Color online) Excitation spectra of $^{122}$Xe resulting from the 
IBM calculation with a single boson-number configuration $[n=9]$ (left panel),
and including configuration mixing between $[n=8]$, $[n=9]$, and  $[n=10]$
 boson spaces (\pv-IBM). Experimental states are from Refs.~\cite{data,garrett2017}
(right panel). The $0^+$ band-head states of the
 $K^\pi=0^+$ bands are highlighted with thick lines.} 
\label{fig:spectra}
\end{center}
\end{figure*}

%-----------------------------------------------------------
%
%	Amplitudes
%
%-----------------------------------------------------------
\begin{figure}[htb!]
\begin{center}
\includegraphics[width=0.8\linewidth]{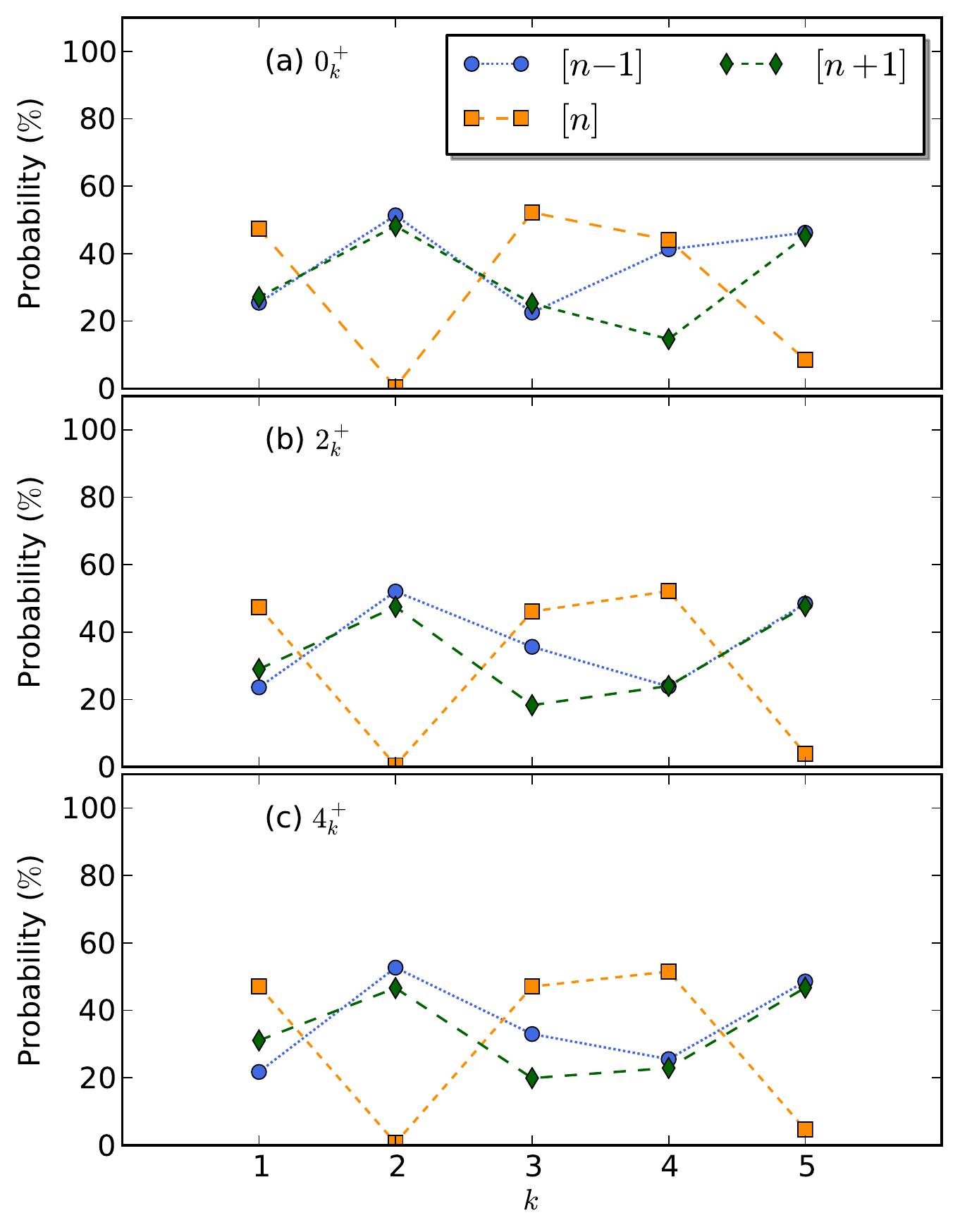} 
\caption{(Color online)  Probabilities of the $[n-1]$, $[n]$, 
and $[n+1]$ components in the wave functions of the five lowest-energy $0^+$, $2^+$, 
and $4^+$ states in $^{122}$Xe. } 
\label{fig:wf}
\end{center}
\end{figure}

%-----------------------------------------------------------
%
%	MIXING MATRIX ELEMENTS
%
%-----------------------------------------------------------
\begin{figure}[htb!]
\begin{center}
\includegraphics[width=0.8\linewidth]{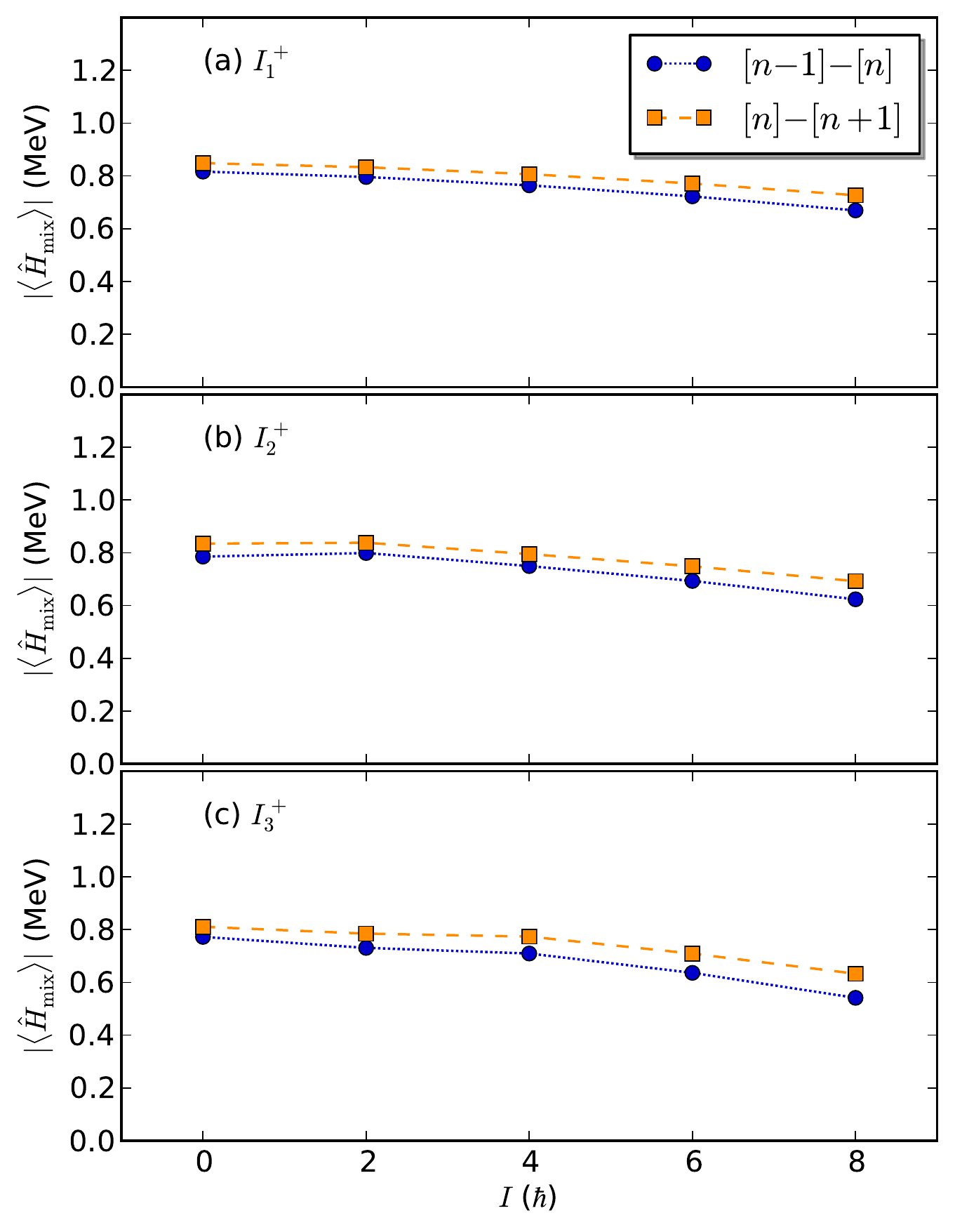} 
\caption{(Color online)  Matrix elements of the monopole pair transfer operator 
$\hat H_\mathrm{mix}$ (last term in Eq.~(\ref{eq:ham2})) between the
 unperturbed $[n-1]$ and $[n]$ configurations, and between the
 unperturbed $[n]$ and $[n+1]$ configurations, for the lowest three
 even-spin states up to $I=8^+$.}
\label{fig:mixmat}
\end{center}
\end{figure}

\section{Illustrative example: $^{122}$Xe\label{sec:xe122}}

\subsection{Energy spectra}

Figure~\ref{fig:unptb} depicts the calculated excitation spectra corresponding to 
the unperturbed boson configurations $[n-1]$, $[n]$ and $[n+1]$, and the spectrum 
obtained by mixing the three different configurations. 
Without mixing, the $0^+$ ground states for the three configurations cluster  
together within a small energy range, and the first excited $0^+$ states are 
also found in a narrow interval around 3 MeV. This is, of course, easy to understand because the 
spaces in which the Hamiltonian is diagonalized only differ by $\Delta n =1$ in the boson number.   
Allowing for configuration mixing (boson-number non-conserving term in the Hamiltonian Eq.~(\ref{eq:ham2})), 
states with the same spin repel and the two lowest $0^+$ excited states are found at
excitation energies $E_\mathrm{exc}\approx 1$ MeV and above $2$ MeV. 
This shows that, using only a single configuration and fixed boson number, the model cannot reproduce 
the excitation energies of low-lying $0^+$ states. Mixing configurations that correspond to different 
boson numbers will be essential for the description of low-energy $0^+$ excitations. 

Figure~\ref{fig:spectra} compares the excitation spectra for $^{122}$Xe
calculated using the IBM with a single configuration $[n]$, 
where the boson number $n=9$ is conserved and the effect of the
pairing vibration is not taken into account (left-hand panel,
(a)), with those obtained with the IBM that includes pairing-vibrations (\pv-IBM), shown in the  
central panel (b)). 
Part of the available experimental energy spectra
\cite{data,garrett2017} is also shown in 
the right-hand panel of Fig.~\ref{fig:spectra}. 
The theoretical states are grouped into bands according to the sequence
of calculated E2 strength values. Since we aim to describe excited $0^+$ states, 
only bands that are built on a $0^+$ state and that
follow the $\Delta I=2$ E2 transition systematics are 
shown in the figure. The remarkable result is that the $K^{\pi}=0^+$ bands 
built on the $0^+_{2}$ and $0^+_3$ states are dramatically lowered in energy by 
taking into account configuration mixing, that is, by the inclusion of 
pairing vibrations. The resulting excitation spectrum is in much better agreement with experiment. 
 
\subsection{Structure of wave functions}

To shed more light upon the nature of excited states calculated in
the \pv-IBM, we show in Fig.~\ref{fig:wf} the probabilities 
of the three different boson-space configurations $[n-1]$, $[n]$, and
$[n+1]$ in the lowest five $0^+$, $2^+$, and $4^+$ states of $^{122}$Xe. 
Let us consider, for example, the $0^+$ states. Only half the wave function 
of the ground state $0^+_1$ is accounted
for by the $[n]$ configuration, while the rest is equally shared by the $[n-1]$ and
$[n+1]$ configurations. 
The $0^+_2$ state exhibits a structure that is completely different from the ground 
state. The dominant contributions come from the $[n-1]$ and $[n+1]$
configurations, both with probabilities of nearly 50 \%,
whereas there is almost no contribution from the $[n]$ configuration space. 
The structure of the $0^+_3$ state is very similar to that of the $0^+_1$. 
The state $0^+_4$ appears to be different from the lower ones 
in that the three configurations are more equally mixed: the
$[n]$ and $[n-1]$ components are found with approximately 40 \% probability each, 
and the remaining 20 \% belongs to the $[n+1]$ configuration space. 
The content of the $0^+_5$ wave function is similar to that of $0^+_2$. 
A corresponding structure is also found for the $2^+$ and $4^+$ states. 
The only exception is perhaps the fourth lowest state of $2^+$ and $4^+$, 
nevertheless in each state $0^+_4$, $2^+_4$, and $4^+_4$ the largest contribution 
to their wave function comes from the $[n]$ configuration.  

\subsection{Mixing matrix elements}

Figure~\ref{fig:mixmat} displays the matrix elements of the mixing
interaction $| \braket{I^+_k  | \hat H_\mathrm{mix} | I^+_k} |$ ($I$
even, $k=1,2,3$), with $\hat H_\mathrm{mix}= t_s(s^\+ + s)$, that couple
%the lowest three even-$I$ states in 
the unperturbed $[n-1]$ and $[n]$ configurations, and the unperturbed
$[n]$ and $[n+1]$ configurations. 
For all of the unperturbed $I_{1,2,3}$ states, the mixing between 
the $[n-1]$ and $[n]$ configurations is almost identical to the  
coupling between the $[n]$ and $[n+1]$ configurations. 
In both cases the mixing is generally stronger between states with lower spin, and 
gradually decreases in magnitude as the angular momentum increases. 

\subsection{Electromagnetic transitions}

The electric quadrupole (E2) and monopole (E0) transition rates can 
also be analyzed in the \pv-IBM. The corresponding operators are defined as 
\begin{align}
& \hat T^{E2}  = e_\mathrm{B}\hat Q \\
\label{eq:e0}
&\hat T^{E0}  = \xi\hat n_d + \eta\hat n
\end{align}
with $e_\mathrm{B}$ is the E2 boson effective charge,  
and $\xi$ and $\eta$ are parameters. 
The $B(E2)$ and $\rho^2(E0)$ transition rates are then calculated using the relations: 
\begin{align}
& B(E2; I_i\to I'_j) = \frac{1}{2I_i+1} | \braket{I'_j \| \hat T^{E2} \| I_i} |^2 \\
& \rho^2(E0; I_i\to I_j) = \frac{Z^2}{e^2r_0^4A^{4/3}} \frac{1}{2I_i + 1}| \braket{I_j \| \hat T^{E0} \| I_i} |^2 \;.
\end{align}

%-----------------------------------------------------------------------
%
%	B(E2)) Xe-122
%
%-----------------------------------------------------------------------
\begin{table}[!htb]
\begin{center}
\caption{\label{tab:e2-xe122} $B({E2}; I_i\to I'_j)$
 values in Weisskopf units, calculated in the IBM and 
\pv-IBM. The experimental values are taken from ENSDF database. }
 \begin{ruledtabular}
 \begin{tabular}{lccc}
  & IBM & \pv-IBM & Experiment \\
\hline
$B({{E2}};2^{+}_{1}\to 0^{+}_{1})$ & 80 & 79 & 78(4) \\
$B({{E2}};4^{+}_{1}\to 2^{+}_{1})$ & 113 & 114 & 114(6) \\
$B({{E2}};6^{+}_{1}\to 4^{+}_{1})$ & 121 & 124 & 1.1$\times 10^2$(4)\\
$B({{E2}};2^{+}_{K=0^+_2}\to 0^{+}_{K=0^+_2})$ & 46 & 79 &  \\ 
% 2+(3) state in IBM
$B({{E2}};4^{+}_{K=0^+_2}\to 2^{+}_{K=0^+_2})$ & 57 & 111 &  \\ % 4+(4) state in IBM
$B({{E2}};6^{+}_{K=0^+_2}\to 4^{+}_{K=0^+_2})$ & 64 & 119 &  \\ % 6+(4) state in IBM
%$\rho^2({E0};0^+_2\to 0^+_1)$ & 125 & 133 & \\
%$\rho^2({E0};0^+_3\to 0^+_1)$ & 122 & 0.05 & \\
%$\rho^2({E0};0^+_3\to 0^+_2)$ & 16 & 120 & \\
%$\rho^2({E0};0^+_4\to 0^+_1)$ & 1.1 & 106 & \\
%$\rho^2({E0};0^+_4\to 0^+_2)$ & 256 & 3.0 & \\
%$\rho^2({E0};0^+_4\to 0^+_3)$ & 66 & 0.01 & \\
 \end{tabular}
 \end{ruledtabular}
\end{center} 
\end{table}

%-----------------------------------------------------------------------
%
%	E0 RME Xe-122
%
%-----------------------------------------------------------------------
\begin{table}[!htb]
\begin{center}
\caption{\label{tab:e0-xe122} Reduced matrix elements of the $d$-boson number
 operator $\hat n_d$ and the total boson number operator $\hat n$ for E0 transitions
 between the lowest three IBM and \pv-IBM states $0^+$ and $2^+$ of $^{122}$Xe.}
 \begin{ruledtabular}
\begin{tabular}{ccccc}
\multirow{2}{*}{} & & IBM & \multicolumn{2}{c}{\pv-IBM} \\
\cline{3-3}
\cline{4-5}
$I^+_i$ & $I^+_j$ & $\braket{I_j \|\hat n_d \| I_i}$ & $\braket{I_j\|\hat n_d\|I_i}$ & $\braket{I_j\|\hat n\|I_i}$ \\
\hline
$0^+_2$ & $0^+_1$ & $-1.166$ & 0.473 & 0.721 \\
$0^+_3$ & $0^+_1$ & $-1.139$ & 0.006 & 0.017 \\
$0^+_3$ & $0^+_2$ & $-0.431$ & 0.451 & 0.687 \\
$2^+_2$ & $2^+_1$ & $-0.503$ & $-0.947$ & $-1.609$ \\
$2^+_3$ & $2^+_1$ & 2.714 & $-0.482$ & 0.084 \\ 
$2^+_3$ & $2^+_2$ & 0.752 & $-0.173$ & $-0.017$ \\
\end{tabular}
 \end{ruledtabular}
\end{center} 
\end{table}

In Table~\ref{tab:e2-xe122} we compare the $B(E2)$ values 
calculated with (\pv-IBM) and without (IBM) the inclusion of 
dynamical pairing. 
A typical value for the E2 effective charge $e_\mathrm{B} = 0.11$
$e\cdot$b is used both in the IBM and \pv-IBM calculations. 
The $B(E2)$ transitions between the
yrast states do not change by the inclusion of the pairing degree of
freedom. The results of both calculations are consistent with the
experimental values \cite{data}. In the \pv-IBM calculation, the E2
transitions in the $0^+_2$-based band display a more pronounced collectivity,
comparable to that in the ground state band. 
%The $B(E2)$ transition strengths within the $K^\pi=0^+_2$ band
%calculated with the \pv-IBM model are about by a factor of two larger
%than those with the IBM (without pairing vibration). 
As shown in Fig.~\ref{fig:wf}, in the \pv-IBM wave functions we find a rather large
contribution from the $[n+1]$ configurations to the $0^+_2$ band, 
and this accounts for the enhanced $B(E2)$ strengths within this 
sequence of states.

Since there are no data for the $E0$ transitions in 
$^{122}$Xe, in Table~\ref{tab:e0-xe122} we compare the calculated
reduced matrix elements of the $\hat n_d$ and $\hat n$ operators,
which constitute the E0 operator of Eq.~(\ref{eq:e0}). 
Note that in the number-conserving IBM only the $\hat n_d$ term
contributes. 
From Table~\ref{tab:e0-xe122} one notices that the reduced matrix elements
$\braket{I_j\|\hat n_d\|I_i}$ in the \pv-IBM calculation are 
systematically smaller in magnitude than the corresponding quantity in
the IBM, most notably for the $0^+_3\to 0^+_1$ transition. 
%This is easily explained by the completely different structure of the
%$0^+_1$ and $0^+_3$ wave functions in the \pv-IBM (see Fig.~\ref{fig:wf}(a)). 
The matrix element $\braket{I_j\|\hat n\|I_i}$ is
generally of equal magnitude as that of $\hat n_d$ and, therefore, one expects 
that it will give a sizeable contribution to the $\rho^2(E0)$
values in \pv-IBM.

\section{Application to $N=92$ isotones\label{sec:N92}}

%-----------------------------------------------------------
%
%	SCMF PES FOR N=92
%
%-----------------------------------------------------------
\begin{figure}[htb!]
\begin{center}
\includegraphics[width=\linewidth]{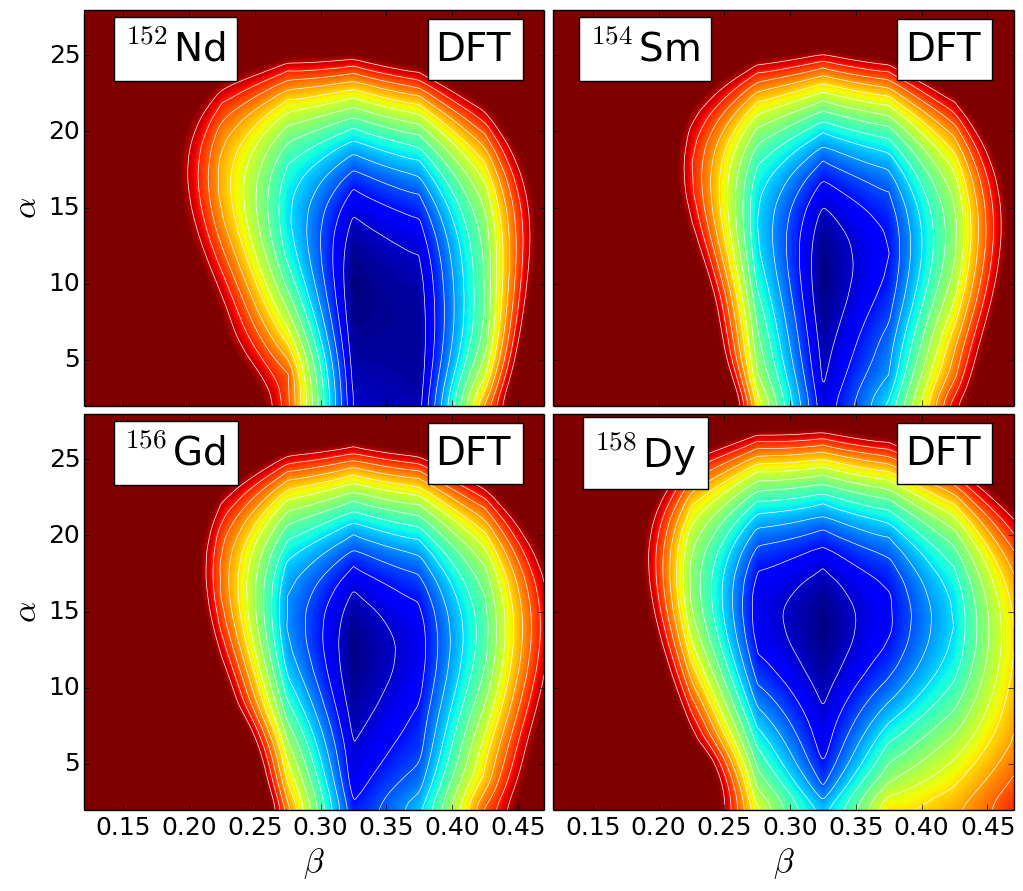}
\caption{(Color online) Same as in the caption to Fig.~\ref{fig:pes-dft}
 but for the $N=92$ isotones $^{152}$Nd, $^{154}$Sm, $^{156}$Gd, and
 $^{158}$Dy. }
\label{fig:pes-dft-N92}
\end{center}
\end{figure}

%-----------------------------------------------------------
%
%	IBM PES FOR N=92
%
%-----------------------------------------------------------
\begin{figure}[htb!]
\begin{center}
\includegraphics[width=\linewidth]{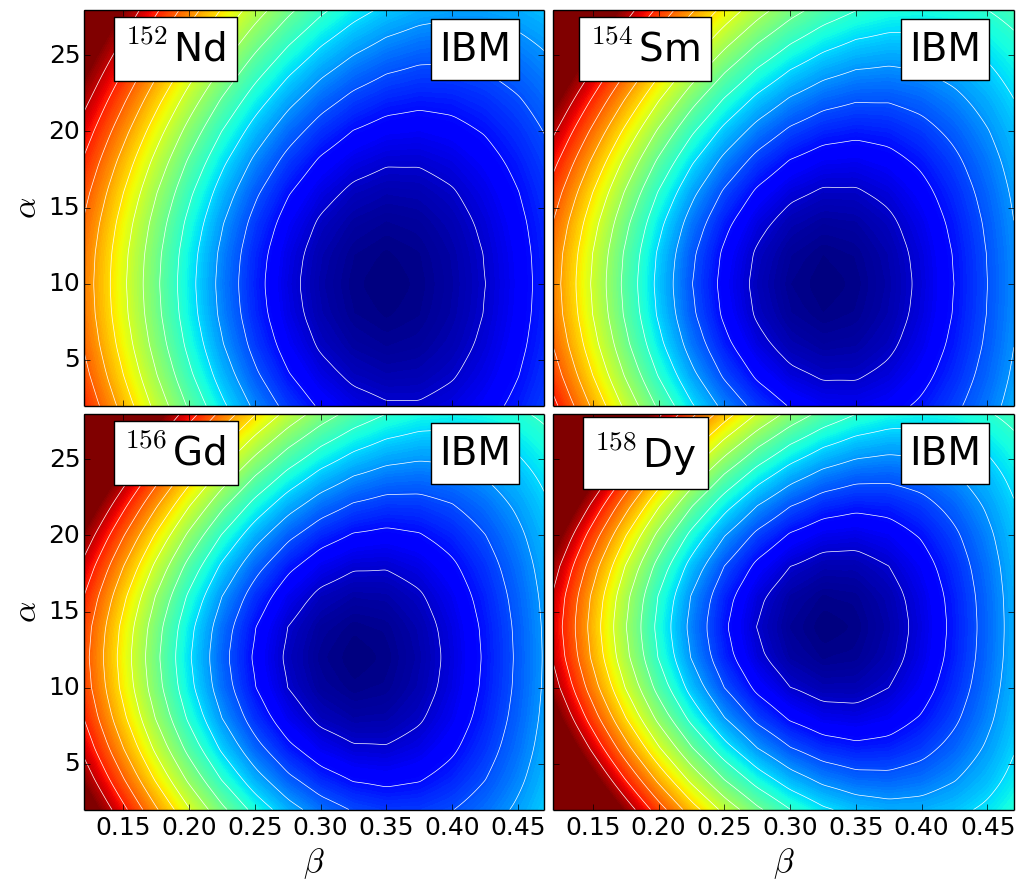} 
\caption{(Color online) Same as in the caption to
 Fig.~\ref{fig:pes-dft-N92}, but for 
 the IBM energy surfaces.}
\label{fig:pes-ibm-N92}
\end{center}
\end{figure}

For a more detailed analysis, we apply the \pv-IBM theoretical framework to 
a study of the structure of the axially-symmetric
$N=92$ rare-earth isotones. For nuclei in this
region of the nuclear chart, an unexpectedly large number of low-energy
excited $0^+$ states have been observed
\cite{aprahamian2018,majola2019}. From a 
theoretical point of view, they have been interpreted in terms of 
pairing vibrations \cite{xiang2020}, contributions of intruder
orbitals \cite{vanisacker1982}, and excitations of double octupole
phonons \cite{zamfir2002,nomura2015}. The occurrence of low-lying 
excited $0^+$ states also characterizes the quantum
shape-phase transition from spherical to axially-deformed
nuclear systems \cite{cejnar2010}.

\subsection{$(\beta,\alpha)$ potential energy surfaces}

In Fig.~\ref{fig:pes-dft-N92} we plot the SCMF
deformation energy surfaces in the $(\beta, \alpha)$ plane for the $N=92$
isotones: $^{152}$Nd, $^{154}$Sm, $^{156}$Gd, and $^{158}$Dy. Note that 
this is the same as Fig. 7 in Ref.~\cite{xiang2020}, in which the coupling of 
shape and pairing vibrations was analyzed using a collective Hamiltonian based 
on nuclear DFT. Pronounced axially symmetric global minima are calculated 
at $\beta \approx 0.35$. The deformation surfaces are much softer with respect 
to the pairing deformation $\alpha$, and the minima extend in a rather large interval 
$5\leqslant\alpha\leqslant 15$. As already noted in Ref.~\cite{xiang2020}, 
this softness is reduced with the increase of
the proton number, while simultaneously the energy surfaces
become more soft in the quadrupole collective deformation.
%-----------------------------------------------------------------------
%
%	PARAMETERS FOR N=92
%
%-----------------------------------------------------------------------
\begin{table}[!htb]
\begin{center}
\caption{\label{tab:para-N92} 
Same as the caption Table~\ref{tab:parameter}, but for the 
$N=92$ isotones. }
 \begin{ruledtabular}
 \begin{tabular}{lcccccccc}
  & $\epsilon_s^0$ & $\epsilon_d$ & $\kappa$ & $\chi$ & $\kappa'$
  &  $t_s$  & $C_\beta$ & $C_\alpha$ \\
\hline
$^{152}$Nd & 1.40 & 0.478 & $-0.045$ & $-1.1$ & $-0.0130$ & 0.16 & 2.85 & 0.035 \\
$^{154}$Sm & 1.37 & 0.626 & $-0.043$ & $-1.1$ & $-0.0126$ & 0.16 & 2.90 & 0.040 \\
$^{156}$Gd & 1.30 & 0.530 & $-0.040$ & $-0.9$ & $-0.0083$ & 0.14 & 2.80 & 0.045 \\
$^{158}$Dy & 1.32 & 0.533 & $-0.038$ & $-0.85$ & $-0.0054$ & 0.12 & 2.75 & 0.050 \\
 \end{tabular}
 \end{ruledtabular}
\end{center} 
\end{table}

The corresponding bosonic energy surfaces in the $(\beta,\alpha)$ plane are drawn in Fig.~\ref{fig:pes-ibm-N92}. 
They exhibit a non-zero $\alpha$ global minimum, consistent with the microscopic SCMF PESs.
As already noted above in the case of $^{122}$Xe, the IBM PESs are considerably softer
than the SCMF ones, especially far from the global minimum. This is due to
the more restricted boson model space, that is, the restricted space of valence nucleons from which the bosons are built 
does not contain the high-energy configurations that contribute to the SCMF solutions far from the equilibrium minimum. 
The strength parameters of the boson Hamiltonian in Eq.~(\ref{eq:ham}), determined by mapping the SCMF energy surfaces 
to the expectation values of the Hamiltonian in the boson condensate, 
are listed in Table~\ref{tab:para-N92} for the $N=92$ isotones . 
The large negative values of the derived parameter $\chi$ parameter, close to the
SU(3) limit of the IBM $\chi_\mathrm{SU(3)}=-\sqrt{7}/2$, reflect the pronounced 
axially-symmetric prolate quadrupole deformation of these nuclei.

%-----------------------------------------------------------
%
%	ND152 SPECTRA
%
%-----------------------------------------------------------
\begin{figure*}[htb!]
\begin{center}
\includegraphics[width=0.8\linewidth]{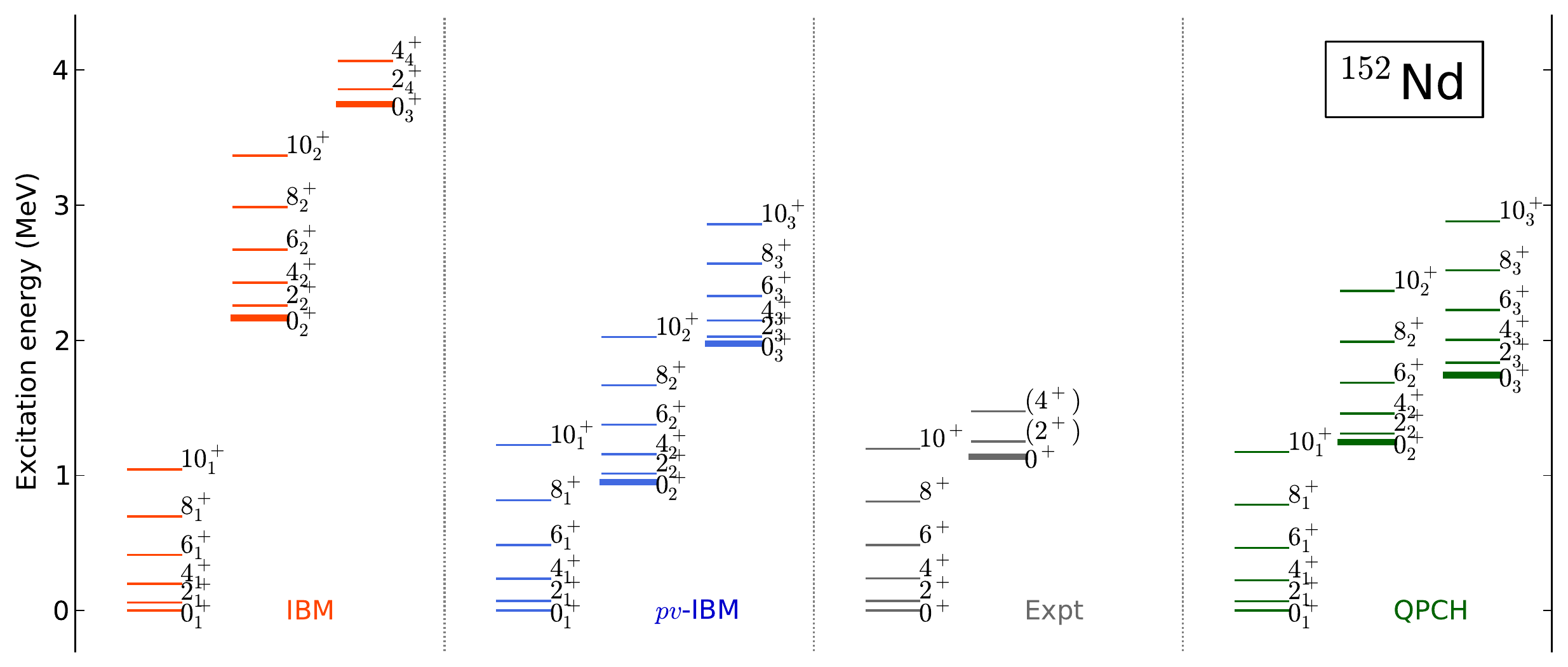} 
\caption{(Color online) Low-energy $K^\pi=0^+$ bands of $^{152}$Nd, 
 calculated using the IBM without and with the dynamical pairing degree
 of freedom, in comparison to available data
 \cite{data}. The corresponding spectrum obtained with the Quadrupole-Pairing
 Collective Hamiltonian (QPCH) model is included for comparison.} 
\label{fig:nd152}
\end{center}
\end{figure*}

%-----------------------------------------------------------
%
%	SM154 SPECTRA
%
%-----------------------------------------------------------
\begin{figure*}[htb!]
\begin{center}
\includegraphics[width=0.8\linewidth]{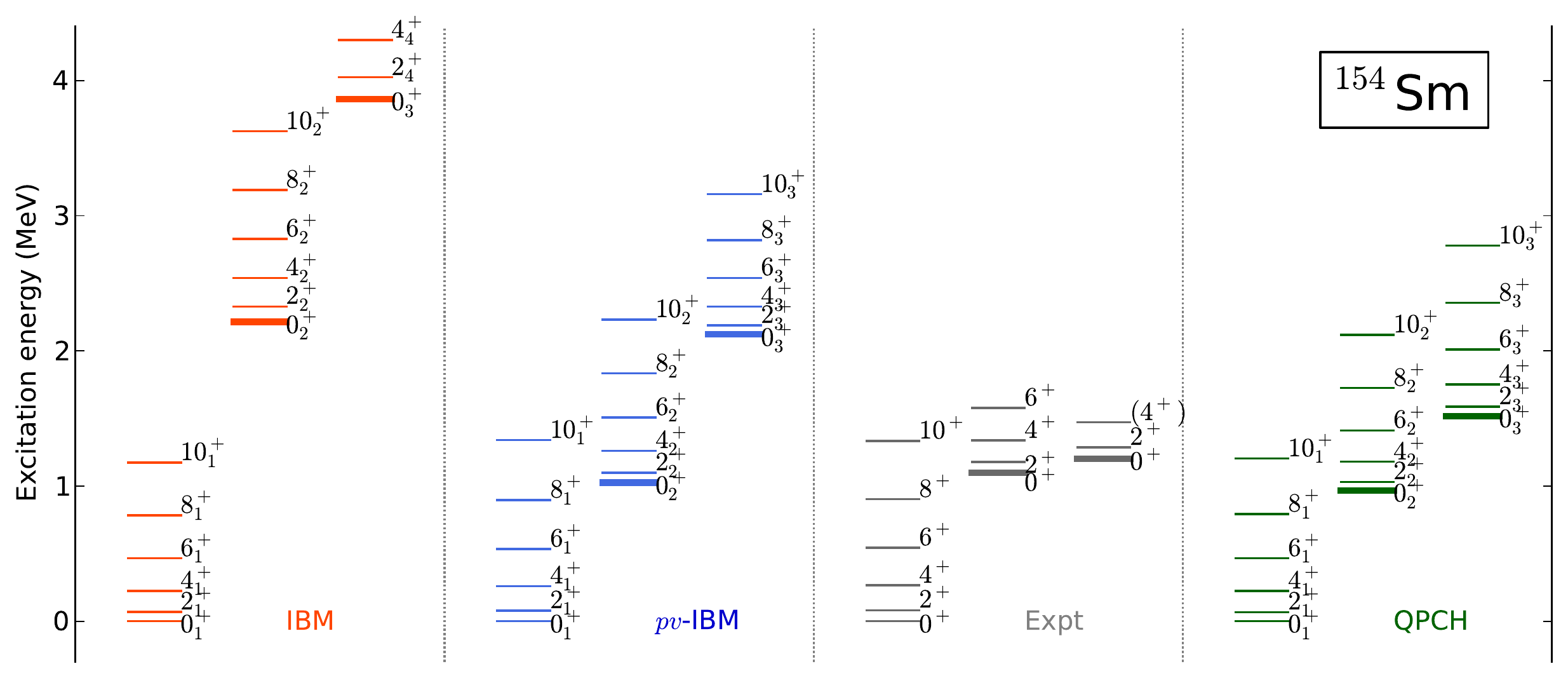} 
\caption{(Color online) Same as in the caption to Fig.~\ref{fig:nd152}, but for $^{154}$Sm.} 
\label{fig:sm154}
\end{center}
\end{figure*}

%-----------------------------------------------------------
%
%	GD156 SPECTRA
%
%-----------------------------------------------------------
\begin{figure*}[htb!]
\begin{center}
\includegraphics[width=0.8\linewidth]{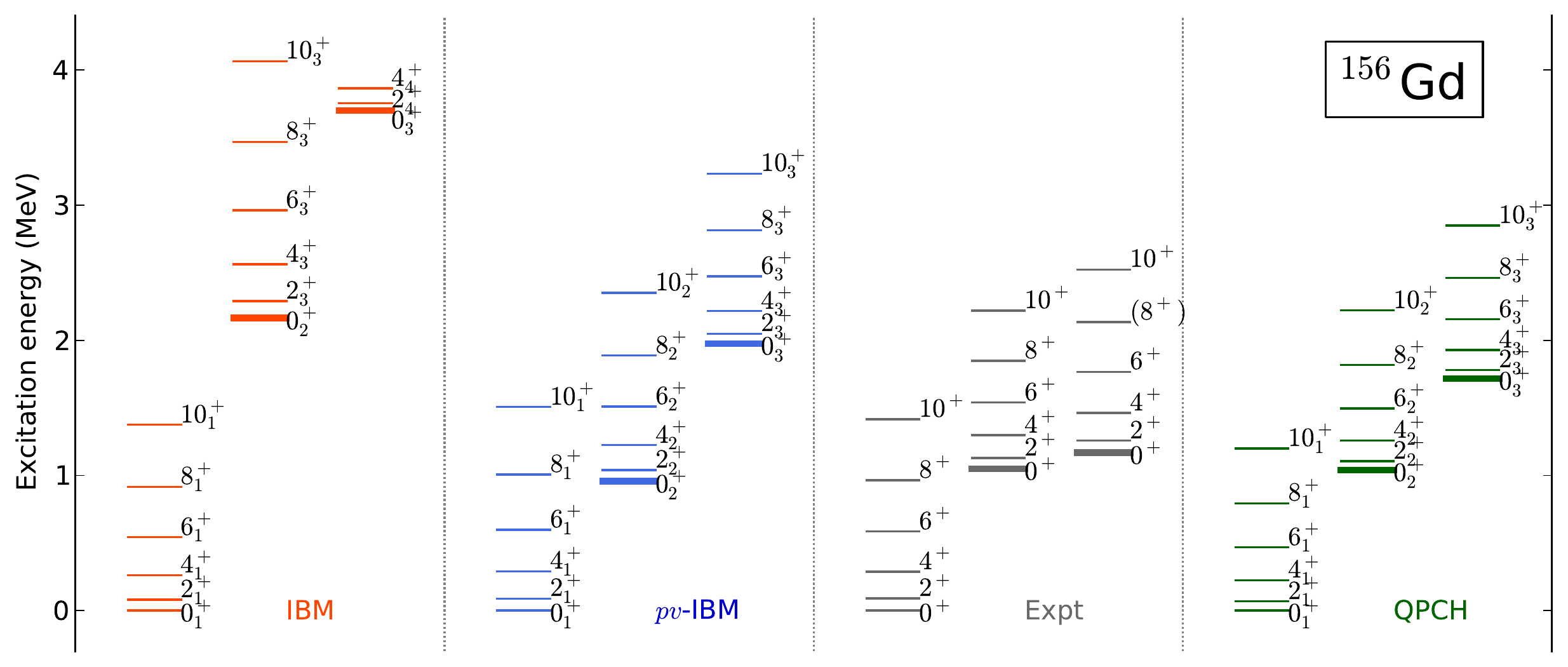} 
\caption{(Color online) Same as in the caption to Fig.~\ref{fig:nd152},
 but for $^{156}$Gd.} 
\label{fig:gd156}
\end{center}
\end{figure*}

%-----------------------------------------------------------
%
%	DY158 SPECTRA
%
%-----------------------------------------------------------
\begin{figure*}[htb!]
\begin{center}
\includegraphics[width=0.8\linewidth]{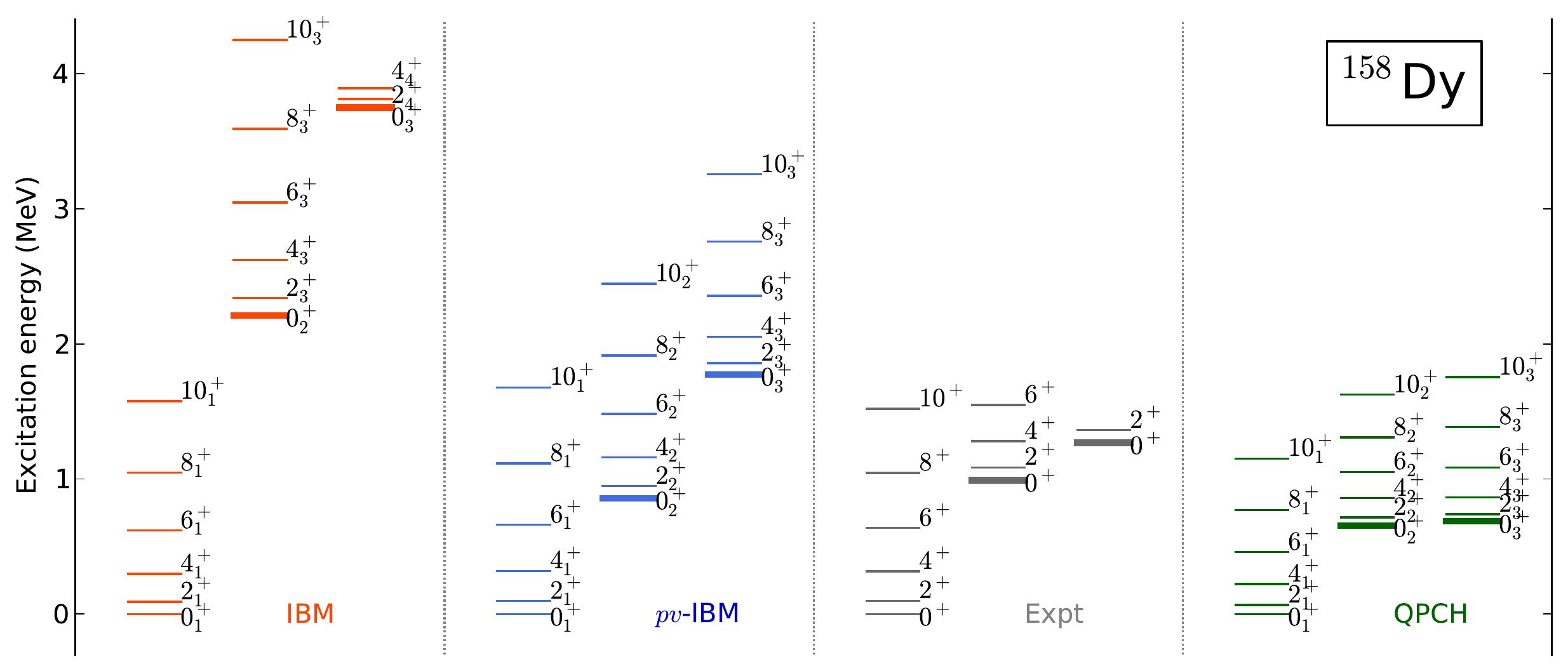} 
\caption{(Color online) Same as in the caption to Fig.~\ref{fig:nd152}, but for $^{158}$Dy.} 
\label{fig:dy158}
\end{center}
\end{figure*}

%-----------------------------------------------------------
%
%	N=92: 0+ WAVE FUNCTIONS
%
%-----------------------------------------------------------
\begin{figure}[htb!]
\begin{center}
\includegraphics[width=\linewidth]{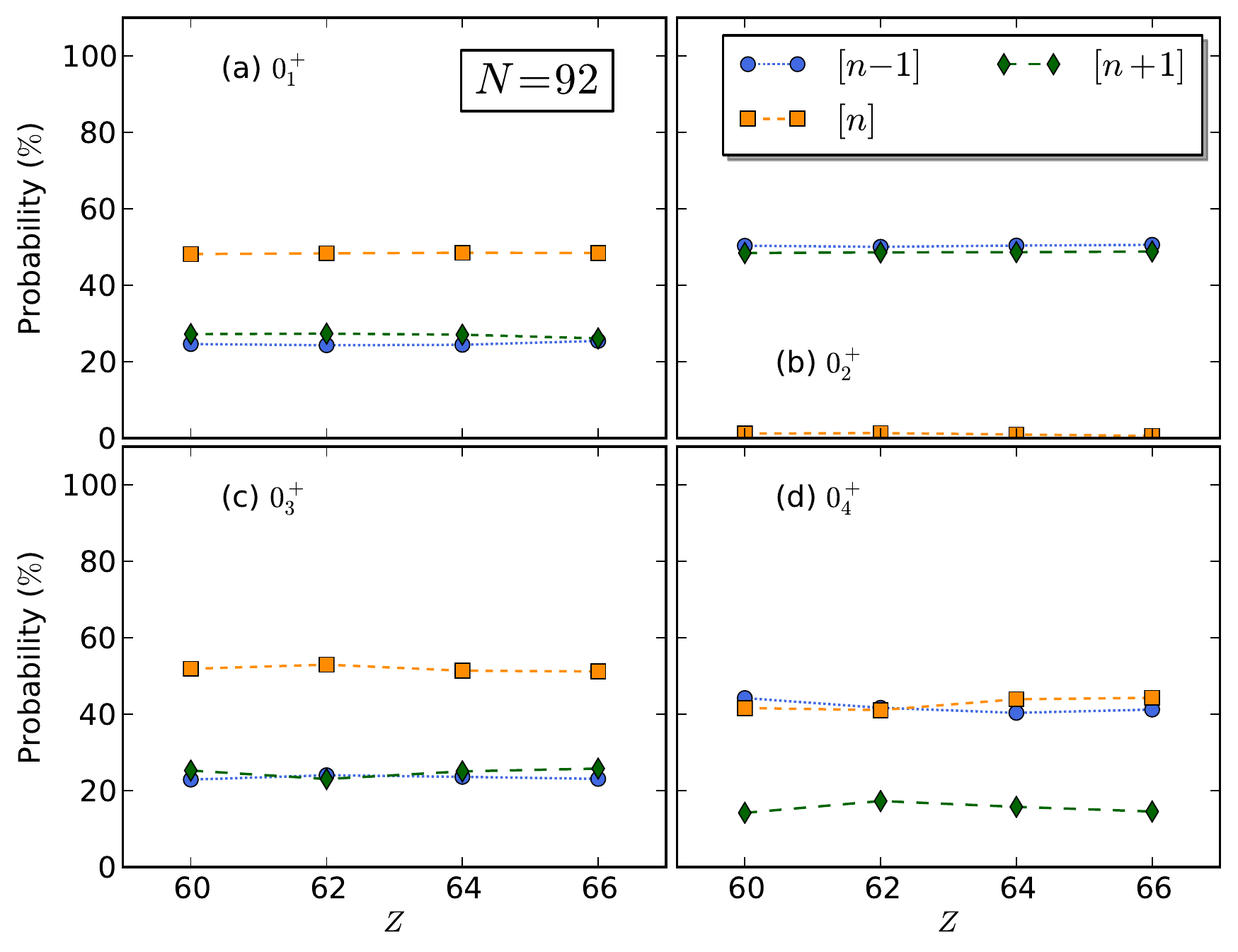} 
\caption{(Color online) Probabilities of the $[n-1]$, $[n]$, 
and $[n+1]$ components in the \pv-IBM wave functions of the four lowest $0^+$ states 
in the $N=92$ isotones.} 
\label{fig:amp_N92}
\end{center}
\end{figure}

\subsection{Low-energy excitation spectra}

Figures~\ref{fig:nd152}, \ref{fig:sm154}, \ref{fig:gd156}, and
\ref{fig:dy158} compare the three-lowest $K^\pi=0^+$ bands of four
$N=92$ isotones: $^{152}$Nd, $^{154}$Sm, $^{156}$Gd, and $^{158}$Dy,
respectively, computed using the IBM and \pv-IBM and IBM. In addition to 
the corresponding data, we also include the results of our recent study that 
has used the newly developed Quadrupole-Pairing Collective Hamiltonian (QPCH) 
to analyze the low-energy spectra of these nuclei \cite{xiang2020}. 
A detailed description of the QPCH model can be found Ref.~\cite{xiang2020}. 
All three Hamiltonians (IBM, \pv-IBM, and QPCH) used here are based on the 
same energy density functional and pairing interaction. 
The excitation spectra shown in Figs.~\ref{fig:nd152}-\ref{fig:dy158} clearly illustrate the 
striking effect of the coupling between shape and pairing degrees of freedom. The 
inclusion of dynamical pairing significantly lowers the bands based on excited $0^+$ states. 
The bands calculated with \pv-IBM and QPCH are in much better agreement with 
experiment, especially the band based on $0^+_2$. We note that the overall quality 
of the \pv-IBM description of $K^\pi=0^+$ bands is comparable to that of the fully microscopic 
QPCH model. 

Even though we only show the $K^\pi=0^+$ bands in 
Figs.~\ref{fig:nd152}-\ref{fig:dy158}, the $K^\pi=2^+$ (or
$\gamma$-) bands are also observed experimentally for the $N=92$
isotones. The IBM models can be used to compute these
states but, since this study is restricted to axial symmetry, the focus is on
$K^\pi=0^+$ bands. For completeness, the $K=2^+_1$ bandhead is
calculated to be 2.248 (2.315), 2.330 (2.451), 2.102 (2.114), and 2.085 (2.099) 
MeV, for $^{152}$Nd, $^{154}$Sm, $^{156}$Gd, and $^{158}$Dy in the
\pv-IBM (IBM) calculations, respectively. 
Thus, in the axial case, the energies of the $\gamma$ band are hardly affected by
the inclusion of the pairing degree of freedom. 
The corresponding experimental $2^+_\gamma$ energies for 
$^{154}$Sm, $^{156}$Gd, and $^{158}$Dy are: 1.440 \cite{smallcombe2014},
1.154 \cite{aprahamian2018}, 0.946 MeV \cite{majola2019}, respectively,
whereas no $\gamma$ band has been identified in $^{152}$Nd. Therefore we note that, 
for a quantitative comparison with data, the theoretical framework should be extended with 
the $\gamma$ degree of freedom (non-axial shapes).

\subsection{Structure of the wave functions}

In Fig.~\ref{fig:amp_N92} we plot the probabilities  
of the three different boson configurations $[n-1]$, $[n]$, 
and $[n+1]$ in the \pv-IBM 
wave functions of the four lowest-energy $0^+$ states. In all four  
$N=92$ isotones nearly half of the wave function of the $0^+_1$ ground state (a) 
is accounted for by the $[n]$ configuration. 
The structure of wave function for the $0^+_2$ state is based mainly on
the $[n-1]$ and $[n+1]$ configurations, with almost no contribution from the 
states of the $[n]$-boson model space. 
The $0^+_3$ state is mainly composed of $[n]$-boson 
configurations, similar to the $0^+_1$ ground state. 
The wave function of the $0^+_4$ state (d) somewhat differs in structure
from the lower-energy $0^+$ states: each of the $[n]$ and $[n-1]$
configurations takes approximately 40 \% of the wave function, and the
remaining 20 \% consists of the $[n+1]$ configuration.

%-----------------------------------------------------------------------
%
%	E.M. TRANSITIONS TABLE N=92 isotones
%
%-----------------------------------------------------------------------
\begin{table}[!htb]
\begin{center}
\caption{\label{tab:e2_N92} $B({E2}; I_i\to I'_j)$ values 
 (in Weisskopf units) for $^{152}$Nd, $^{154}$Sm, $^{156}$Gd, and
 $^{158}$Dy. Experimental values \cite{data,moeller2012,aprahamian2018} are compared to 
 results of the \pv-IBM, IBM and QPCH model calculations.} 
 \begin{ruledtabular}
 \begin{tabular}{llcccc}
 &  & Expt & \pv-IBM & IBM & QPCH \\
\hline
% Nd-152
$^{152}$Nd
& $2^{+}_{1}\to 0^{+}_{1}$ & 173$\pm 10$ & 160 & 162 & 162 \\
& $4^{+}_{1}\to 2^{+}_{1}$ & 226$\pm 11$ & 225 & 227 & 231 \\
& $6^{+}_{1}\to 4^{+}_{1}$ & 218$^{+51}_{-35}$ & 240 & 241 & 253 \\
% Sm-154
$^{154}$Sm
& $2^{+}_{1}\to 0^{+}_{1}$ & 176$\pm 1$ & 184 & 186 & 197 \\
& $4^{+}_{1}\to 2^{+}_{1}$ & 245$\pm 6$ & 260 & 262 & 282 \\
& $6^{+}_{1}\to 4^{+}_{1}$ & 289$\pm 8$ & 280 & 281 & 310 \\
& $8^{+}_{1}\to 6^{+}_{1}$ & 319$\pm 17$ & 283 & 281 & 324 \\
& $0^{+}_{K=0^+_2}\to 2^{+}_{1}$ & 11.2$\pm 2.1$ & 6.0 & 5.0 & 5.9 \\
& $2^{+}_{K=0^+_2}\to 0^{+}_{1}$ & 0.32$\pm 0.04$ & 0.7 & 1.7 & 1.1 \\
& $2^{+}_{K=0^+_2}\to 2^{+}_{1}$ & 0.72$\pm 0.09$ & 1.4 & 2.0 & 1.6 \\
& $2^{+}_{K=0^+_2}\to 4^{+}_{1}$ & 1.32$\pm 0.15$ & 3.7 & 0.6 & 3.1 \\
& $4^{+}_{K=0^+_2}\to 2^{+}_{1}$ & 0.32$\pm 0.11$ & 0.7 & 0.9 & 1.5 \\
& $4^{+}_{K=0^+_2}\to 4^{+}_{1}$ & 0.57$\pm 0.18$ & 1.2 & 1.3 & 1.4 \\
& $4^{+}_{K=0^+_2}\to 6^{+}_{1}$ & 0.66$\pm 0.21$ & 3.7 & 1.9 & 2.7 \\
& $2^{+}_{\gamma}\to 0^{+}_{1}$ & 1.9$\pm 0.2$ & 0.1 & 0.3 & \\ % expt. g-band on the 2+(4) state
& $2^{+}_{\gamma}\to 2^{+}_{1}$ & 3.2$\pm 0.3$ & 2.4 & 1.8 & \\
& $2^{+}_{\gamma}\to 4^{+}_{1}$ & 0.36$\pm 0.05$ & 2.4 & 2.8 & \\
% Gd-156
$^{156}$Gd 
& $2^{+}_{1}\to 0^{+}_{1}$ & 189$\pm 3$ & 195 & 197 & 205 \\
& $4^{+}_{1}\to 2^{+}_{1}$ & 264$\pm 4$ & 276 & 279 & 293 \\
& $6^{+}_{1}\to 4^{+}_{1}$ & 295$\pm 8$ & 299 & 300 & 322 \\
& $8^{+}_{1}\to 6^{+}_{1}$ & 320$\pm 17$ & 304 & 303 & 335 \\
& $10^{+}_{1}\to 8^{+}_{1}$ & 314$\pm 14$ & 299 & 296 & 342 \\
& $0^{+}_{K=0^+_2}\to 2^{+}_{1}$ & 8$^{+4}_{-7}$ & 4.9 & 4.3 & 2.5 \\
& $2^{+}_{K=0^+_2}\to 0^{+}_{1}$ & 0.63$\pm 0.06$ & 0.6 & 0.8 & 1.8 \\
& $2^{+}_{K=0^+_2}\to 0^{+}_{K=0^+_2}$ & 52$\pm 23$ & 196 & 136 & 236 \\
& $2^{+}_{K=0^+_2}\to 2^{+}_{1}$ & 3.3$\pm 0.3$ & 1.2 & 0.5 & 2.5 \\
& $2^{+}_{K=0^+_2}\to 4^{+}_{1}$ & 4.1$\pm 0.4$ & 3.0 & 2.9 & 4.2 \\
& $4^{+}_{K=0^+_2}\to 2^{+}_{K=0^+_2}$ & 330$^{+110}_{-130}$ & 276 & 180
		  & 337 \\
& $4^{+}_{K=0^+_2}\to 2^{+}_{1}$ & 1.3$^{+0.5}_{-0.7}$ & 0.7 & 0.8 &
		      2.6 \\
& $4^{+}_{K=0^+_2}\to 6^{+}_{1}$ & 2.1$^{+0.7}_{-1.1}$ & 3.0 & 3.3 &
		      3.4 \\
& $0^{+}_{K=0^+_3}\to 2^{+}_{1}$ & 1.6$^{+2.3}_{-0.8}$ & 0.02 & 0.02 & 2.6 \\
& $2^{+}_{\gamma}\to 0^{+}_{1}$ & 4.68$\pm 0.16$ & 2.4 & 2.4 & \\ % exp g-band on the 2+(2) state
& $2^{+}_{\gamma}\to 2^{+}_{1}$ & 7.24$\pm 0.25$ & 6.3 & 5.8 & \\
& $2^{+}_{\gamma}\to 4^{+}_{1}$ & 0.77$\pm 0.04$ & 0.2 & 0.2 & \\
%
% Dy-158
$^{158}$Dy 
& $2^{+}_{1}\to 0^{+}_{1}$ & 186$\pm 4$ & 218 & 220 & 199 \\
& $4^{+}_{1}\to 2^{+}_{1}$ & 266$\pm 15$ & 309 & 311 & 284 \\
& $6^{+}_{1}\to 4^{+}_{1}$ & 3.4$\times 10^2(4)$ & 335 & 337 & 312 \\
& $8^{+}_{1}\to 6^{+}_{1}$ & 3.4$\times 10^2(7)$ & 342 & 342 & 326 \\
% exp g-band on the 2+(2) state
& $2^{+}_{\gamma}\to 0^{+}_{1}$ & 5.9$\pm 1.2$ & 3.1 & 3.1 & \\ 
& $2^{+}_{\gamma}\to 2^{+}_{1}$ & 19$\pm 4$ & 7.2 & 6.7 & \\
& $2^{+}_{\gamma}\to 4^{+}_{1}$ & 2.1$\pm 0.8$ & 0.3 & 0.3 & \\
% expt. K=0+(2) band on the 2+(3) state
& $2^{+}_{K=0^+_2}\to 0^{+}_{1}$ & 2.1$\pm 0.5$ & 0.6 & 0.8 & 1.6 \\
& $2^{+}_{K=0^+_2}\to 2^{+}_{1}$ & 3.5$\pm 0.8$ & 1.1 & 0.6 & 1.8 \\ 
& $2^{+}_{K=0^+_2}\to 4^{+}_{1}$ & 12$\pm 3$ & 2.8 & 2.9 & 2.0 \\
 \end{tabular}
 \end{ruledtabular}
\end{center} 
\end{table}

%-----------------------------------------------------------------------
%
%	\rhoE0 for, Sm-154, Gd-156, Dy-158
%
%-----------------------------------------------------------------------
\begin{table}[!htb]
\begin{center}
\caption{\label{tab:e0_N92} Same as the caption to
 Tab.~\ref{tab:e2_N92} but for the $\rho^2({E0}; I_i\to
 I_j)\times 10^3$ values. The experimental $\rho^2(E0)$ are from
 Refs.~\cite{data,backlin1982,wood1999,kibedi2005,wimmer2009,smallcombe2014,majola2019}.}
 \begin{ruledtabular}
 \begin{tabular}{llcccc}
  & & Expt & \pv-IBM & IBM & QPCH \\
\hline
% Sm-154
$^{154}$Sm 
& $0^+_2\to 0^+_1$ & 96$\pm 42$ & 43 & 39 & 54 \\ % Wimmer 2009
& $2^+_2\to 2^+_1$ & $\leqslant 9.4\pm 1.5$ & 41 & 36 & 53 \\ % PLB732, 161 (2014)
& $4^+_2\to 4^+_1$ & 8.2$^{+12.0}_{-8.2}$ & 38 & 28 & 53 \\ % PLB732,161 (2014)
% Gd-156
$^{156}$Gd 
& $0^+_2\to 0^+_1$ & 42$\pm 21$ & 42 & 43 & 73 \\
& $0^+_3\to 0^+_1$ & 1.2$^{+1.9}_{-0.6}$ & 0.2 & 1.8 & 13 \\
& $0^+_3\to 0^+_2$ & 18$^{+27}_{-9}$ & 41 & 40 & 97 \\ % ENSDF
& $0^+_4\to 0^+_1$ & 2.9$^{+2.7}_{-1.4}$ & 54 & 0.07 & 3.4 \\ %kibedi2005
& $0^+_4\to 0^+_3$ & 6.3$^{+5.7}_{-3.0}$ & 0.6 & 5.6 & 34 \\ %kibedi2005
& $2^+_{K=0^+_2}\to 2^+_1$ & 54$\pm 4$ & 40 & 41 & 72 \\
& $2^+_{K=0^+_3}\to 2^+_1$ & 0.2$^{+0.6}_{-0.2}$ & 0.05 & 0.4 & 13 \\% Wood1999
& $4^+_{K=0^+_2}\to 4^+_1$ & 50$^{+25}_{-16}$ & 38 & 34 & 72 \\ % Backlin 1982
& $4^+_{K=0^+_3}\to 4^+_1$ & $<15$ & 4$\times 10^{-5}$ & 0.08 & 13 \\ % Backlin 1982
% Dy-158
$^{158}$Dy
& $2^+_{K=0^+_2}\to 2^+_1$ & 27$\pm 12$ & 40 & 48 & 75 \\ % Wood 1999
 \end{tabular}
 \end{ruledtabular}
\end{center} 
\end{table}

\subsection{Transition rates}

The $B(E2)$ and $\rho^2(E0)$ values calculated with the \pv-IBM, IBM and 
QPCH models are compared to available data in 
Tabs.~\ref{tab:e2_N92} and \ref{tab:e0_N92}, respectively,
The effective boson charge in the E2
operator is $e_\mathrm{B}=0.14$ $e\cdot$b. 
The parameters of the E0 operators: $\xi=0.095$ and
$\eta=0.11$ fm$^2$ for the \pv-IBM, and
$\xi=0.075$ fm$^2$ for the IBM, are adjusted to obtain the best agreement 
with the experimental $\rho^2(E0)$ values for $^{156}$Gd, and kept unchanged for
all four $N=92$ isotones. There are no adjustable parameters for the calculation of 
transition rates in the QPCH model. Note that there are no E2 transitions related to the 
$\gamma$ band with $K^{\pi}=2^+$ in the QPCH model since, as pointed out above, the present
version of QPCH does not include the triaxial degree of freedom. 

As shown in Table~\ref{tab:e2_N92}, the $B(E2)$ transition strengths
within the ground state bands are reproduced very nicely by all the
models. There is no significant difference between the $B(E2)$  values
calculated with the IBM and \pv-IBM. 
Many experimental results are available for the transition rates of 
$^{154}$Sm and, generally, they are well reproduced by all three models. 
In $^{156}$Gd the theoretical results reproduce the data, except for an  
overestimate of the experimental $B(E2;2^{+}_{K=0^+_2}\to
0^{+}_{K=0^+_2})$ value of $52\pm 23$ W.u. Very good results are also obtained 
for $^{158}$Dy.

The calculated $\rho^2(E0)$ values are generally in satisfactory agreement 
with available data (Table \ref{tab:e0_N92}), except in the case of 
$^{154}$Sm, in which both the IBM and QPCH approaches considerably
overestimate the measured \cite{smallcombe2014} upper limits of the
$\rho^2({E0};2^+_2\to 2^+_1)$ and $\rho^2({E0};4^+_2\to
4^+_1)$ values. It appears that the IBM and \pv-IBM models reproduce the data 
somewhat better than QPCH, but this comes at the expense of additional 
adjustable parameters in the $E0$ operator. 
%This does not seem to be improved just by playing with
%the E0 parameters: the calculated reduced matrix elements
%$\braket{2^+_2\|\hat n_d\|2^+_1}=-1.34$ and $\braket{2^+_2\|\hat
%n\|2^+_1}=-1.58$ have the same sign with equal magnitude, and it is not
%possible to reproduce at the same time the $\rho^2({E0};0^+_2\to
%0^+_1)$ value since,
%$\braket{0^+_2\|\hat n_d\|0^+_1}=-0.61$ and $\braket{0^+_2\|\hat
%n\|0^+_1}=-0.71$. 

\section{Conclusion and outlook\label{sec:summary}}

We have developed a model that incorporates the coupling 
between nuclear shape and pairing degrees of freedom in the framework of the IBM,
based on nuclear DFT. To account for pairing vibrations, a boson-number
non-conserving IBM Hamiltonian is introduced. 
The boson model space is then extended from the usual one in which the boson number equals half the 
number of valence nucleons, to include three subspaces that
differ in boson number by one. The three subspaces are mixed by a specific monopole pair 
transfer operator. 
In a first step of the construction of the IBM Hamiltonian, a set of constrained SCMF calculation
is performed for a specific choice of the universal EDF and
pairing force, and with the constraints on the expectation values of the axial mass quadrupole 
operator and monopole pairing operator. These calculations produce a potential energy 
surface (PES) in the plane of the axial quadrupole $\beta$ and pairing
$\alpha$ collective coordinates. The energy surface is then mapped onto the expectation value of the 
IBM Hamiltonian in the boson condensate state. The mapping determines the strength 
parameters of the IBM Hamiltonian, and from the corresponding eigenvalue equation excitation energy 
spectra and transition rates are obtained.

As a first application of the newly developed model, this work has focused on the 
excitation spectrum of $^{122}$Xe. By the inclusion of the dynamical pairing degree of freedom 
in the IBM and the resulting boson-number configuration mixing, it has been shown that the excitation
energies of the $0^+_2$ and $0^+_3$ states and the bands built on them, are dramatically lowered 
by a factor of two or three, thus bringing the theoretical spectrum in quantitative agreement with 
experiment.  The validity of the method has been further examined in a more
systematic study of the axially-symmetric $N=92$ rare-earth isotones. The microscopic coupling 
between shape and pairing degrees of freedom leads 
to a boson Hamiltonian that, when compared to the standard IBM, significantly lowers the 
$K^\pi=0^+$ bands based on excited $0^+$ states in $^{152}$Nd,
$^{154}$Sm, $^{156}$Gd, and $^{158}$Dy. The calculated excitation
spectra are in an excellent agreement with experiment, and are fully 
consistent with the results of the corresponding 
quadrupole-pairing collective Hamiltonian model \cite{xiang2020}. 
Both models also reproduce the empirical  E2 and E0 transition properties with a
reasonable accuracy.

The present study has shown a new interesting possibility for extending the 
DFT-to-IBM mapping method. By incorporating explicitly the dynamical
pairing degree of freedom in the IBM, this model can be used to describe 
pairing vibrational modes and quantitatively reproduce the excitations
of low-energy $0^+$ states. Here we have only considered the coupling 
of the pairing degree of freedom with the axial shape deformation. 
 A more challenging case, but also more realistic, will be the 
coupling between the pairing and triaxial $(\beta,\gamma)$ shape degrees
of freedom. This will be particularly important in
$\gamma$-soft nuclei and systems that exhibit shape coexistence. In principle, 
such an extension is also possible in the QPCH approach, however this 
generates additional terms in the collective Schr\"odinger
equation that represent the couplings of the $\beta-\gamma$ and
$\gamma-\alpha$ variables. In contrast, it is rather straightforward to
extend the present IBM framework to triaxial nuclei, since there
is no need for new building blocks in the boson Hamiltonian. Work in 
this direction is in progress, and will be reported in a forthcoming
article.

\begin{acknowledgments}
This work has been supported by the Tenure Track Pilot Programme of 
the Croatian Science Foundation and the 
\'Ecole Polytechnique F\'ed\'erale de Lausanne, and 
the Project TTP-2018-07-3554 Exotic Nuclear Structure and Dynamics, 
with funds of the Croatian-Swiss Research Programme. 
It has also been supported in part by the QuantiXLie Centre of Excellence, a project co-financed by the Croatian Government and European Union through the European Regional Development Fund - the Competitiveness and Cohesion Operational Programme (KK.01.1.1.01).
\end{acknowledgments}

\bibliography{refs}

\end{document}